\documentclass[twocolumn,showpacs,preprintnumbers,amsmath,amssymb,superscriptaddress]{revtex4-1} 
\usepackage{chemformula} 
\usepackage{graphicx}
\usepackage[colorlinks=true,linkcolor=blue,citecolor=blue]{hyperref}
\usepackage{color}

\begin{document}

\title{Directing random lasing emission using cavity exciton-polaritons}

\author{P. Bouteyre}
\affiliation{Universit\'e Paris-Saclay, ENS Paris-Saclay, CNRS, Centrale Supelec, LuMIn, 91405 Orsay, France}
\author{H. S. Nguyen}
\affiliation{Universit\'e de Lyon, Insitut des Nanotechnologies de Lyon - INL, UMR CNRS 5270, CNRS, Ecole Centrale de Lyon, Ecully, F-69134, France}
\author{J.-S. Lauret}
\affiliation{Universit\'e Paris-Saclay, ENS Paris-Saclay, CNRS, Centrale Supelec, LuMIn, 91405 Orsay, France}
\author{G. Tripp\'e-Allard}
\affiliation{Universit\'e Paris-Saclay, ENS Paris-Saclay, CNRS, Centrale Supelec, LuMIn, 91405 Orsay, France}
\author{G. Delport}
\affiliation{Universit\'e Paris-Saclay, ENS Paris-Saclay, CNRS, Centrale Supelec, LuMIn, 91405 Orsay, France}
\author{F. L\'ed\'ee}
\affiliation{Universit\'e Paris-Saclay, ENS Paris-Saclay, CNRS, Centrale Supelec, LuMIn, 91405 Orsay, France}
\author{H. Diab}
\affiliation{Universit\'e Paris-Saclay, ENS Paris-Saclay, CNRS, Centrale Supelec, LuMIn, 91405 Orsay, France}
\author{A. Belarouci}
\affiliation{Universit\'e de Lyon, Insitut des Nanotechnologies de Lyon - INL, UMR CNRS 5270, CNRS, Ecole Centrale de Lyon, Ecully, F-69134, France}
\author{C. Seassal}
\affiliation{Universit\'e de Lyon, Insitut des Nanotechnologies de Lyon - INL, UMR CNRS 5270, CNRS, Ecole Centrale de Lyon, Ecully, F-69134, France}
\author{D. Garrot}
\affiliation{Groupe d$^\prime$Etude de la Mati\`ere Condens\'ee, Universit\'e de Versailles Saint Quentin En
Yvelines, Universit\'e Paris-Saclay, 45 Avenue des Etats-Unis, 78035, Versailles, France}
\author{F. Bretenaker}
\affiliation{Universit\'e Paris-Saclay, ENS Paris-Saclay, CNRS, Centrale Supelec, LuMIn, 91405 Orsay, France}
\author{E. Deleporte }
\email{emmanuelle.deleporte@ens-paris-saclay.fr}
\affiliation{Universit\'e Paris-Saclay, ENS Paris-Saclay, CNRS, Centrale Supelec, LuMIn, 91405 Orsay, France}

\date{\today}

\begin{abstract}
Random lasing \cite{Wiersma2008,Cao2005} is an intriguing phenomenon occurring in disordered structures with optical gain. In such lasers, the scattering of light provides the necessary feedback for lasing action \cite{letokhov1967stimulated,Wiersma1996,Cao1999}. Because of the light scattering, the random lasing systems emit in all the directions \cite{Cao1999b} in contrast with the directional emission of the conventional lasers. While this property can be desired in some cases, the control of the emission directionality remains required for most of the applications. Besides, it is well known that the excitation of cavity exciton-polaritons is intrinsically directional \cite{Weisbuch1992,Panzarini1999,Sanvitto2016a}. Each wavelength (energy) of the cavity polariton, which is a superposition of an excitonic state and a cavity mode, corresponds to a well defined propagation direction. We demonstrate in this article that coupling the emission of a 2D random laser with a cavity polaritonic resonance permits to control the direction of emission of the random laser. This results in a directional random lasing whose emission angle with respect to the microcavity axis can be tuned in a large range of angles by varying the cavity detuning. The emission angles reached experimentally in this work are 15.8$^\circ$ and 22.4$^\circ$. 
\end{abstract}


\maketitle
In a random laser, cavity feedback is replaced by multiple scattering \cite{Wiersma2008,Cao2005,letokhov1967stimulated,Wiersma1996,Cao1999}. For example, in the case of coherent feedback random lasing, light is scattered in closed loops and forms random cavities \cite{Cao1999,Cao1998}. The lasing emission is then characterized by narrow peaks, similar to the conventional lasers, originating from the random cavity modes. In contrast with conventional lasers, random lasers emit in all directions \cite{Cao1999b}. Different methods have been proposed and/or demonstrated for controlling the random lasing directionality, such as pump shaping \cite{Hisch2013}, coupling the random lasing medium to a Bragg grating \cite{Song2009}, to an optical fiber \cite{Turitsyn2010}, to a photonic crystal microcavity \cite{Long2006} or to a planar microcavity \cite{Song2006,Song2007,Song2009a,Ni2015,Schonhuber2016}. Controlling the shape and size of a perturbated photonic crystal is also a way to tame the lasing properties, including the emission wavelength and the mode area, which also leads to a control of a directionality \cite{Lee2019}. All these techniques are based on the coupling of the random laser emission with a more directional resonance mechanism. Typically, the random lasing occurs in two dimensions, and the third dimension is used to sandwich the random laser inside a directional output coupling system. However, there exists another system in which the emission direction is governed by the coupling between a light cavity mode and an excitonic resonance, namely the cavity exciton-polariton resonance \cite{Weisbuch1992,Panzarini1999,Sanvitto2016a}. In this case, the strong coupling between light and matter leads to modes, which are superposition states of light and matter. This results in the existence of two cavity polaritonic branches for which the emission direction is correlated with the energy of the quasi-particle. One could thus imagine coupling a 2D random laser with a cavity polaritonic resonance in order to control the emission direction of the random laser.   \\

\vspace{-5pt}

To implement such an approach, one needs to find a material in which both random lasing and strong coupling between excitons and cavity modes have been obtained. This is the case of organic-inorganic hybrid perovskites. Lasing with perovskites has been demonstrated with several different resonators in the past such as nanowires \cite{zhu2015lead}, nanoplatelets \cite{zhang2014room}, distributed feedback cavities (DFB) \cite{saliba2016structured}, photonic crystals \cite{chen2016photonic} and  microcavities \cite{deschler2014high}. Additionally, an interesting feature of perovskites is the continuous tunability of the emission wavelength through the entire visible spectrum via halide substitution \cite{Kim2014}. Moreover, in the present context, the advantage of these materials is that random lasing can also be observed without an external resonator, i.e. directly from polycrystalline and nanocrystals thin films. This has indeed been demonstrated in iodine-based perovskites \cite{kao2014lasing,dhanker2014random,Shi2016,Kao2016a,Safdar2018}, chloride-based perovskites \cite{yakunin2015low} and bromide-based perovskites \cite{yakunin2015low,Li2017a,Li2017,Yuan2018a,Xu2018,Liu2016,Roy2018,Weng2018,Fan2019,Mikosch2019,Weng2019,Wang2019,Tang2019,Liu2019}. Besides, strong coupling between a microcavity mode and an exciton has been observed both in 2D and 3D hybrid perovskites \cite{Brehier2006,Lanty2008b,Su2017,Su2018,Bouteyre2019}.  \\

\vspace{-5pt}

In this letter, we investigate random lasing action with coherent feedback in a polycrystalline thin film of the perovskite CH$_3$NH$_3$PbBr$_3$ capped with PMMA. This 2D random laser is embedded in a microcavity in which  strong coupling between the cavity mode and the perovskite exciton is obtained, and we demonstrate that coupling the emission of this 2D random laser with the cavity polaritonic resonance permits to control the direction of emission of the random laser.  

\begin{figure}[h!]
\centering
  \includegraphics[width=\linewidth]{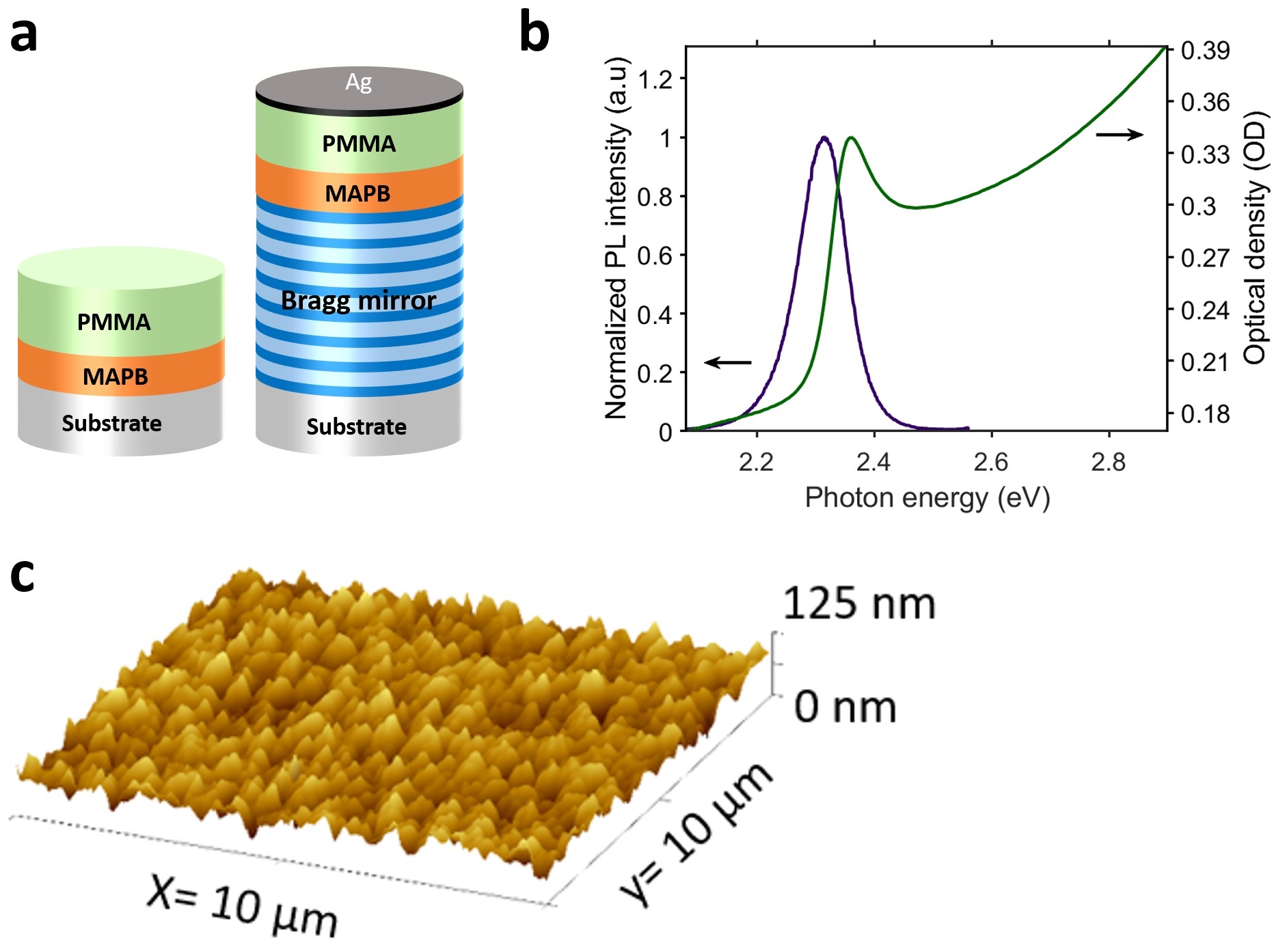}
  \caption{\textbf{a)} Sketches of the two samples \textbf{b)} Photoluminescence (PL) and absorption spectra of the MAPB thin film. \textbf{c)} AFM image of the MAPB thin film}
 \label{Figure1}
\end{figure}

Two samples, sketched in figure \ref{Figure1} a), have been studied in this work. The first sample is a 100 nm-thick thin film of the perovskite CH$_3$NH$_3$PbBr$_3$, called hereafter MAPB, deposited on a quartz substrate by spin coating. A layer of around 350 nm of PMMA (Poly(methyl) methacrylate) has been later deposited by spin coating on the MAPB layer. The second sample is a 3$\lambda/2$ MAPB-based microcavity, with $\lambda=535$ nm the MAPB emission wavelength. This microcavity is composed by a commercial Bragg mirror (Layertec, corp) on which have been deposited the same two layers of MAPB ($\sim$ 100nm) and PMMA ($\sim$ 350 nm) by spin coating and a layer of 30 nm of silver by evaporation. The two samples fabrication is further detailed in the section 1 of the supplementary. 
The figure \ref{Figure1} b) shows the absorption and photoluminescence spectra of a control thin film of MAPB. The absorption spectrum is composed by an excitonic resonance at around 2.35 eV and a band absorption continuum at higher energies. The PL is Stokes-shifted at an energy of 2.32 eV with a Full Width at Half Maximum (FWHM) of 96 meV. Figure \ref{Figure1} c) presents an AFM scan of a control layer of MAPB: it reveals a polycrystalline film with a roughness of about 15 nm (RMS(grain-wise)=14.29 nm) composed by grains of 500 nm to 1 $\mu$m size. Photoluminescence (PL) spectroscopy and angle-resolved photoluminescence (ARPL) as a function of the pump power have been performed on both samples. More details on the optical set-ups are given in the methods section of this article. 

Due to the roughness of the MAPB layer, random lasing action is observed from the MAPB/PMMA sample. Such random lasing from perovskite polycrystalline thin films have already been demonstrated in the literature in MAPB thin layers \cite{Wang2019,Liu2019}. The PL spectroscopy demonstrating the random lasing is shown in the section 2 of the supplementary. Figures \ref{Figure2} a) and b) show the ARPL pseudo-colour maps below (at 0.4 $P_{th}$) and above (at 1.3 $P_{th}$) the random lasing threshold obtained on a position of the MAPB/PMMA sample. More data on this position such as the ARPL maps for other pumping powers are given in the section 3 of the supplementary. The ARPL map below the excitation threshold shows an angle-independent emission centred at 2.3 eV with an FWHM of 100 meV. Above the threshold, two random laser peaks can be clearly observed as two horizontal red lines at 2.275 eV and 2.283 eV while the broadband PL signal similar to the one below the threshold appears in dark blue. Such as the PL emission below the threshold, the random laser peaks emit in all directions from the sample's surface as expected for a random lasing action in thin films \cite{Cao1999b}. Indeed, the gain occurs within the plane of the MAPB film and the lasing emission is scattered out of the plane, resulting in an angle-independent emission from the surface of the MAPB/PMMA sample.

\begin{figure}[!ht]
\centering
\includegraphics[width=\linewidth]{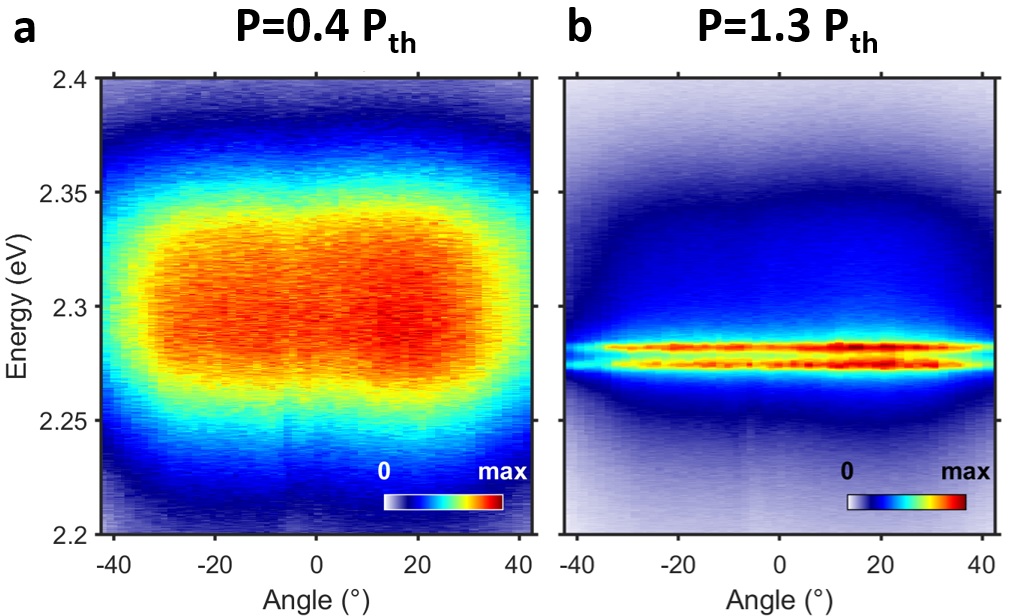}
  \caption{Angle-resolved photoluminescence (ARPL) pseudo-colour maps (in linear scale), a) under (0.4 $P_{th}$) and b) above (1.3 $P_{th}$) the random lasing threshold of the thin film of MAPB on quartz substrate capped with PMMA. The resolutions of the ARPL maps are respectively 1.7 meV for the energy axis and 0.7$^\circ$ for the angle axis.}
  \label{Figure2}
\end{figure}

Considering the ease with which the strong coupling regime is observed in MAPB \cite{Bouteyre2019,Zhang2017,Shang2018a}, the question of the photonic or polaritonic nature of the lasing can be reasonably raised here as it was raised by Niyuki et al. from a resonance-controlled ZnO random laser \cite{Niyuki2017}. Especially, the blueshift of the MAPB/PMMA lasing peaks (of around 1-2 meV between 1.1 and 1.8 $P_{th}$, see section 2 of the Supplementary) could be in favour of a polaritonic lasing. However, the blueshift could just be explained by thermal effects as MAPB shows blueshift with temperature increase (of around 24 meV between 300K and 370K) \cite{Wright2016}. Nevertheless, to distinguish between the photonic and polaritonic nature of the lasing is hard and not within our reach in this paper. Indeed, the only way to demonstrate without ambiguity the existence of the polariton is to measure the mode dispersion from the random lasing emission and it is impossible in practice here due to the multiple light scattering. \\

Let us now focus on the 3$\lambda$/2 MAPB-based microcavity. In our previous study \cite{Bouteyre2019}, the strong coupling between the perovskite's exciton and the microcavity photonic mode has been demonstrated on a similar microcavity and cavity polaritons have been observed experimentally (more details on the theory of the cavity polaritons are given in the methods section). In particular, we have shown that, due to the overall lateral roughness of the MAPB and PMMA layers, the cavity detuning, $\delta=E_0-E_X$, with $E_0$ the cavity mode energy at normal incidence and $E_X$ the exciton energy, can be tuned by probing different lateral positions. This was due to the overall lateral roughness of the MAPB and PMMA layers. Only the lower cavity polariton dispersion can be observed in the ARPL maps of our previous study \cite{Bouteyre2019}. We will then only consider the lower cavity polariton whose eingenvalue $\mu_{LP}(\theta)$ is given by equation \ref{eq1} :


\begin{equation}
\label{eq1}
\centering
\begin{split}
&\mu_{LP}(\theta)=\frac{1}{2}[E_{ph}(\theta)+E_{X}-i(\gamma_{ph}+\gamma_{X})]\\
&-\sqrt{V^2+\frac{1}{4}[E_{X}-E_{ph}(\theta)+i(\gamma_{ph}-\gamma_{X})]^2},\\
& \\
&\text{with} \quad E_{ph}(\theta)=\frac{\delta+E_X}{\sqrt{1-\frac{sin^2(\theta)}{n_{eff}^2}}},
\end{split}
\end{equation}

\noindent where $E_{ph}(\theta)$ is the cavity dispersion, $n_{eff}$ is the cavity effective refractive index, $\gamma_{ph}$ and $\gamma_{X}$ are the linewidths of the photonic mode and the exciton, and V is the coupling strength. The real part of $\mu_{LP}(\theta)$ corresponds to the lower cavity polariton energy dispersion, $E_{LP}(\theta)$, and the imaginary part to the lower cavity polariton linewidth, $\gamma_{LP}(\theta)$. From the study, the values of $n_{eff}$, $E_X$, $\gamma_{ph}$, $\gamma_{X}$ and V were found to be 1.75, 2.355 eV, 25 meV, 90 meV and 48 meV respectively. \\ 

\begin{figure*}[!ht]
\centering
\includegraphics[width=0.8\linewidth]{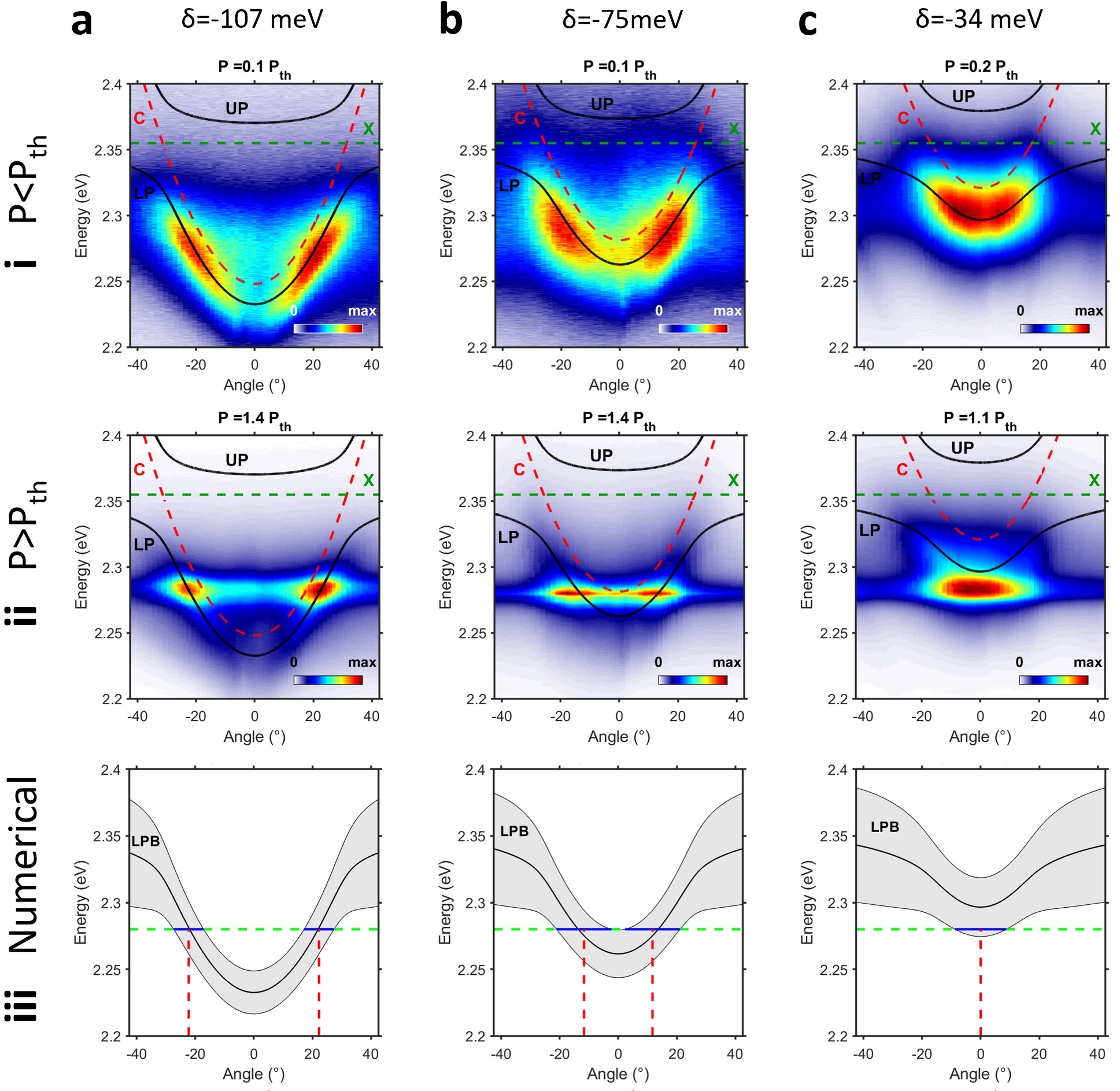}
  \caption{Angle-resolved photoluminescence (ARPL) pseudo-colour maps (in linear scale) of three different positions on the MAPB-based microcavity (a) to c)). The resolutions of the ARPL maps are respectively 1.7 meV for the energy axis and 0.7$^\circ$ for the angle axis. The first row (i) corresponds to the ARPL maps below the lasing threshold and the second row (ii) above the lasing threshold. The lower and upper cavity polaritons dispersions (black lines), cavity mode dispersions (dashed red line) and the exciton energy (green dashed line) are plotted on top of all the ARPL maps. The dispersions are derived from the equations in the methods section with $n_{eff}$=1.75, $E_X$=2.355 eV, $\gamma_{ph}$ =25 meV, $\gamma_{X}$=90 meV and $V$= 48 meV. The detunings are $\delta$=-107 meV, -75 meV and -34 meV, respectively. The figures in the third row (iii) are the numerical lower cavity polaritons dispersions plotted with the lower cavity polariton linewidths (gray shaded area) using the same parameters mentioned above. The centre of the lower cavity polariton dispersions are plotted as black solid lines. The green dashed lines correspond to a laser emission energy at 2.28 eV, the blue solid lines correspond to the intersections between the laser emission energy and the lower cavity polariton dispersions, the red dashed line indicates the expected angle of emission.}
  \label{Figure3}
\end{figure*}

\begin{figure*}[ht!]
\centering
  \includegraphics[width=\linewidth]{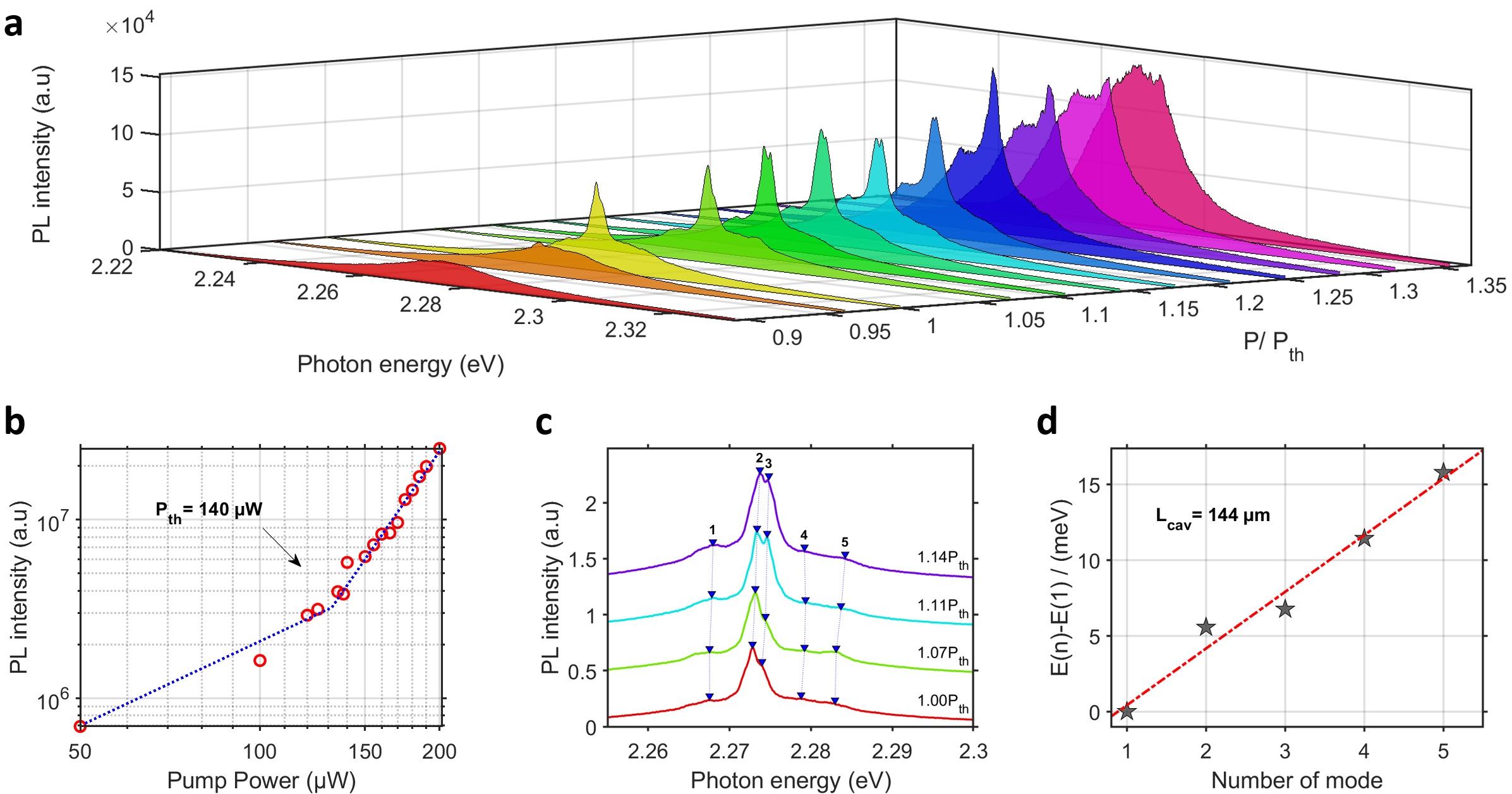}
  \caption{PL spectroscopy of the lasing action from the $3\lambda/2$ microcavity. The resolution is here of 0.38 meV for the energy axis. \textbf{a)} PL spectra of one position of the MAPB-based microcavity at different pumping powers. \textbf{b)} Log-log PL intensity spectrum as a function of the pumping power. \textbf{c)} Four PL spectra above the lasing threshold of a) are plotted with a vertical offset for better readability. Five lasing modes are indicated by blue triangles and the blue dotted lines are guides for the eye. \textbf{d)} Study of the free spectral range in energy, $FSR_E=E_{m+1}-E_{m}$ with $m$ an integer, of the lasing spectra in c) to obtain the characteristic size of the MAPB pseudo-cavities. The difference in energy between the $m^{th}$ modes and the $1^{st}$ mode is plotted against the modes numbers and is fitted with a linear function. The pseudo-cavity characteristic length of 144 $\mu$m is retrieved from $L=hc/(n s)$, where $n=2.3$ is the MAPB refractive index, $c$ is the light velocity and $s$ is the slope of the linear function. } 
 \label{Figure4}
\end{figure*}


The figures \ref{Figure3} a), b) and c) show the ARPL pseudo-colour maps below (in the row (i)) and above (in the row (ii)) the lasing threshold of three different positions on the microcavity. Additional data are shown in the section 3 of the supplementary for each of these positions, including the map-integrated PL intensity as a function of the pump power demonstrating the lasing action of these positions. The dispersions corresponding to the three positions below the threshold (at 0.1, 0.1 and 0.2 $P_{th}$) are fitted with the lower cavity polariton dispersion, with the detuning, $\delta$, as the only free parameter and the other parameters fixed at the values found in our previous study \cite{Bouteyre2019}. More details on the fitting method is presented in the section 4 of the supplementary. The detunings obtained for the three different positions are $\delta$=-120 meV ($E_0$=2.235 eV), $\delta$=-85 meV ($E_0$=2.27 eV) and $\delta$=-35 meV ($E_0$=2.22 eV) respectively. A good agreement is met between the lower cavity polariton dispersions and the dispersions of the ARPL maps confirming the strong coupling regime in this microcavity. The corresponding upper cavity polaritons dispersions, cavity mode dispersions and the exciton energy are also plotted on top of all the ARPL maps (see the methods section).\\

In the three ARPL maps above the threshold (at 1.4, 1.4 and 1.1 $P_{th}$) in the row (ii) of figures \ref{Figure3} a), b) and c), laser peaks appear in red while the lower cavity polariton emissions, similar to the ones below the threshold, appear in cyan and dark blue. To obtain the lasing peaks energies, angles and divergences, vertical and horizontal slices of the ARPL maps are respectively taken at given angles and energies (see the section 5 of the supplementary). For the first position corresponding to the largest detuning in figure \ref{Figure3} a), two laser peaks appear in dark red at 2.282 eV and 2.286 eV at the angles of $\pm$ $\sim$ 22.4$^\circ$ with a divergence of 12.5$^\circ$. For the second position probed on the microcavity (figure \ref{Figure3} b)), one laser peak in red emerges at 2.28 eV at angles around $\pm$ $\sim$ 15.8$^\circ$ with a divergence of $\sim$ 12.3$^\circ$. For these two cases, the angles at which the lasing peaks lie correspond to the intersection between the lower cavity polariton dispersion and the energy of lasing emission. For the third position probed on the microcavity (figure \ref{Figure3} c)), two peaks emerge at 2.282 eV and 2.286 eV in dark red. However, unlike the first two positions, the emission occurs at 0$^\circ$ with a divergence around 29.7$^\circ$. Moreover, the positions of the laser peaks lie under the theoretical lower cavity polariton dispersion. \\


Figure \ref{Figure4} a) shows the photoluminescence spectra as a function of the pump power varying from 0.9 $P_{th}$ to 1.35 $P_{th}$. The photoluminescence spectrum below the threshold is coupled to the microcavity mode, which leads to a spectrum shifted to 2.272 eV with an FWHM of 53 meV. When the pump power exceeds the threshold, laser peaks appear on top of the broadband PL with a dominant peak at 2.273 eV with an FWHM of 3 meV. Such as for the PL spectroscopy of the MAPB/PMMA sample (see section 2 of the supplementary), when the pump power further increases, other peaks occur at different energies, and the overall lasing spectrum is broadened as the modes begin to overlap. The log-log integrated PL as a function of the pumping power of figure \ref{Figure4} b) exhibits a threshold of 140 $\mu$W. Note that the threshold is around two hundred times higher than the MAPB/PMMA sample threshold of 0.65 $\mu$W (see section 2 of the supplementary). This is due to the high absorption of the silver mirror at 405 nm.

One could wonder if the microcavity lasing is the case of a polaritonic lasing which is the condensation of the cavity polaritons. A slight blueshift of the lasing modes, of about 1 meV between 1 $P_{th}$ and 1.15 $P_{th}$, can be observed in figure \ref{Figure4} c)  (showing four lasing spectra of figure \ref{Figure4} a) with a vertical offset for better readability) with the pumping power increasing, but as already discussed previously, can be due to the perovskite thermal effects. A first strong argument to rule out the condensation of cavity polaritons is the fact that an emission at lower energies than the theoretical cavity polaritonic dispersion is seen in figure \ref{Figure3} c) : this observation can not be explained by the cavity polaritonic lasing. However, in the case of the two positions corresponding to large detunings in figures \ref{Figure3} a), and b), the lasing emission at large angles could suggest a propagation of a polariton condensate, having the consequence that the polariton condensate can be observed at non-zero in-plane wavevectors. The polariton condensate propagation was recently observed with perovskites in a microcavity containing microwires of \ch{CsPbBr_3} \cite{Su2018}. This phenomenon is due to the polariton interactions with the excitonic reservoir, which ejects the polariton condensate radially from the center of the pump spot and provide the condensate with kinetic energy \cite{Wouters2008}. The phenomenon only takes place when the pump spot size is smaller than the polariton condensate propagation length. For larger spots, the polariton condensate lie at $k_{//}$=0. For our system, the estimated propagation length of the polariton condensate is around 500 nm (see calculation in Methods). In our experiment, the pump laser beam was focused by a 0.9 NA objective for the first position in figure \ref{Figure3} a) and by a 5 cm lens for the positions in figures \ref{Figure3} b) and c). Thus, we observe that lasing emission occurs at large angles whether the microcavity is pumped with a spot whose radius is smaller or larger than 500 nm, which is a second strong argument to discredit the hypothesis of a propagation of the polariton condensate. As a conclusion, two arguments lead us to rule out the polaritonic lasing. 


\begin{figure*}[ht!]
\centering
 \includegraphics[width=\linewidth]{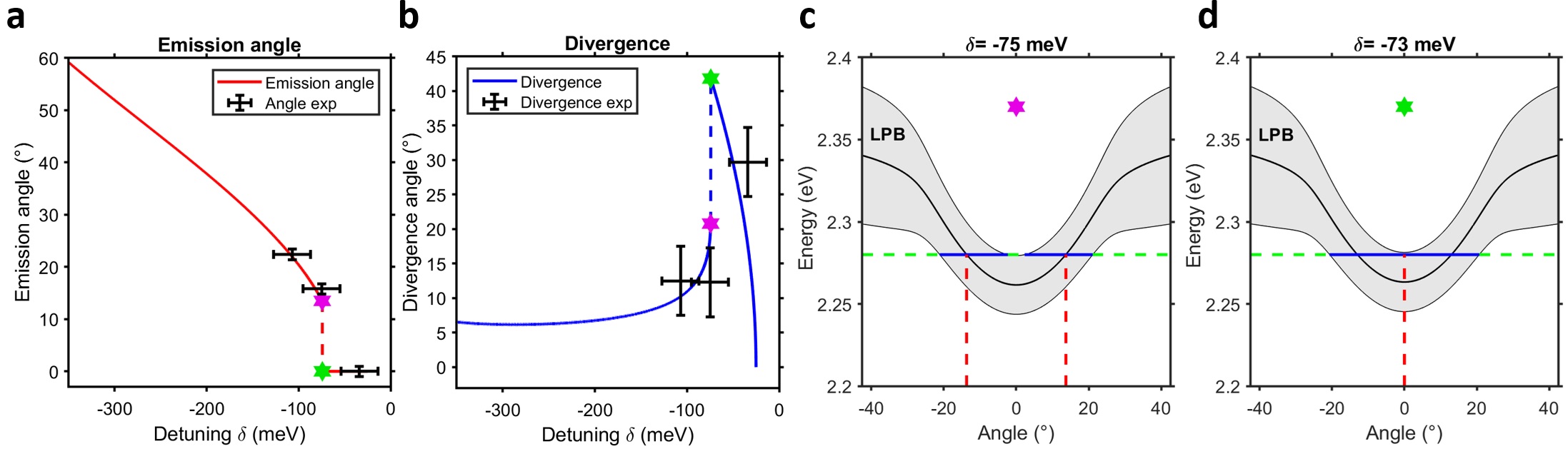}
  \caption{Comparison of the experimental and the numerically expected emission angles \textbf{a)} and divergences \textbf{b)}. The red and blue solid lines correspond to the numerical results of respectively the lasing emission angle and divergence. The black crosses correspond to the experimental results. The purple stars correspond to the case of the detuning $\delta=-75$ meV illustrated in \textbf{c)} and the green stars to the detuning $\delta=-73$ meV in \textbf{d)}. \textbf{c)} and \textbf{d)} show the transition between the two cases of coupling, below and above $\delta=-74 meV$.}
  \label{Figure5}
\end{figure*}

One could also wonder whether the lasing from the microcavity is just simply the case of a photonic laser from a VCSEL. In this case, only one lasing mode would have been expected in the range of photon energy of interest as the length of the microcavity is around 3 times the half of the MAPB emission wavelength (3$\lambda$/2). However, several modes, even though one mode dominates the others, have been observed in figure \ref{Figure4}. The PL spectra on another position of the microcavity (see the section 6 of the supplementary) exhibit also, and more clearly, several modes which confirms the multimodal behavior of the microcavity lasing. This multimodal aspect allows us then to rule out the photonic lasing from the possible types of lasing.\\

On the other hand, we note that the lasing peaks seen in figure \ref{Figure3} and \ref{Figure4} occur at the same range of energy, between 2.26 eV and 2.29 eV, than the MAPB/PMMA sample (See section 2 of the supplementary). Indeed, for a random lasing action, the gain medium properties define the energy region of the lasing emission. Finally, the study in figure \ref{Figure4} d) of the free spectral range in energy, $FSR_E=E_{m+1}-E_{m}$ with $m$ an integer, of the lasing spectra in figure \ref{Figure4} c) gives a characteristic pseudo-cavity length (i.e. the length of the closed loops formed by the multiple scattering) of around 144 $\mu$m (see more details in the methods section), which is similar to the one of 100 $\mu$m found for the MAPB/PMMA sample (See section 2 of the supplementary). A value of the same order of magnitude (of around 215 $\mu$m) has been found for another position of the microcavity in the section 6 of the supplementary. All these arguments are then in favor of the random lasing within the MAPB layer of the microcavity filtered directionally by the lower cavity polariton resonance. As a matter of fact, we think that in ref \cite{Lai2016}, the authors have observed such a filtering on the p-band lasing in a strong coupled ZnO microcavity.\\

To further investigate the coupling of the random lasing with the lower cavity polariton resonance, the expected lasing angle and divergence as a function of the detuning is determined using the equation \ref{eq1} and the parameters found from the fit of the dispersions. The energy of the random lasing emission is considered here to be 2.28 eV. Although the angle of emission can be easily derived in the case of the weak coupling, i.e. V=0 meV, the analytical solution is not trivial in the case of the strong coupling. A numerical approach illustrated by the figures in the row (iii) of figure \ref{Figure3} has been then chosen. The lower cavity polariton dispersions are plotted for a given detuning $\delta$ by taking into account the lower cavity polariton linewidths, $\gamma_{LP}^{\delta}(\theta)$, and are represented by the gray shaded areas. The center of the dispersions, $E_{LP}^{\delta}(\theta)$, are plotted as black solid lines. The intersections between the lasing emission energies (green dashed lines) and the cavity polariton dispersions (gray shaded areas), give then the expected lasing angles and divergences.\\

For large negative detunings, lower than -74 meV, the intersection between the lasing emission energy and the cavity polariton dispersion gives rise to two distinct lobes of lasing emission. The two lobes at negative and positive angles are symmetric with respect to the normal direction. In this case, the two angles of emission (red dashed line) are given by the two intersections between the lasing emission energy (green dashed line) and the center of the cavity polariton dispersion (black solid line). The width in angle of the intersections between the lasing emission energy and the total lower cavity polariton dispersion (gray shaded area) determines the dispersion angle (solid blue line). The ARPL maps obtained from the first two positions of the microcavity in figure \ref{Figure3} a) and b) in the lines (i) and (ii) as well as their theoretical dispersions in the line (iii) clearly demonstrate the cavity polaritonic filtering. When the detunings are above -74 meV, the intersection between the lasing emission energy and the lower cavity polariton dispersion generates only one lasing lobe centered at 0$^\circ$ with a large divergence. This is due to the two lobes merging into one due to their increased widths. This is the case of the emission of the third position probed on the microcavity in figure \ref{Figure3} c). Figures \ref{Figure5} c) and d) illustrating the numerical methods at the detunings $\delta$=-75 and -73 meV shows the transition between the two cases of coupling detailed previously. Figure \ref{Figure5} a) shows the numerical results for the expected emission angle (red line) and figure \ref{Figure5} b) the divergence (blue line) both along with the experimental results (black cross). Only the lasing lobes at positive angles are plotted in figure \ref{Figure5}. The agreement between the experimental and numerical lasing angles and divergences validate the directional filtering of the lasing emission by the lower cavity polariton resonance.\\

In conclusion, we have shown that the directionality of the random lasing in a \ch{CH_3NH_3PbBr_3} polycrystalline thin layer can be controlled by the lower cavity polariton resonance of a 3$\lambda$/2 microcavity in the strong coupling regime. In particular, we highlight here that we have obtained a directional random lasing assisted by the cavity polaritons at angles as large as ± 20$^\circ$ with divergence angles of around 15$^\circ$ and that the angle of emission can be controlled in such a system by changing the detuning, that is to say the thickness, of the microcavity. To decrease the emission divergence, one has to decrease the lower cavity polariton linewidth by improving the microcavity quality factor (see section 7 of the supplementary). Directing random laser emission through a photonic structure acting as an external knob is expected to open bright perspectives in the area of hybrid perovskites optoelectronics. Beyond vertical Fabry-Perot cavities, emission could be filtered by photonic crystals or dispersion-engineered metasurfaces, bringing additional degrees of freedom, together with a high compactness \cite{Nguyen2018}. Used as external wavelength and direction filters rather than laser cavities, such nanophotonic structures can be technologically tolerant, and compatible with electrical injection.

\section*{Methods}
\noindent \textbf{PL and ARPL as a function of the excitation power} \textit{Photoluminescence (PL) spectroscopy and angle-resolved photoluminescence (ARPL) as a function of the excitation power have been performed on the two samples with a femtosecond pulsed laser emitting at 405 nm (150 fs, 1 kHz rep.rate). The ARPL optical set-up is shown in the supplementary of our previous article \cite{Bouteyre2019} and sketches of the different excitation configurations are shown in the section 8 of the supplementary.} \\
\textit{For the MAPB/PMMA sample, the ARPL maps of the figure \ref{Figure2} and the PL spectroscopy in the section 2 of the supplementary were obtained by using a first configuration : a 0.9 NA objective was used for focusing the pump laser and collecting the PL from the PMMA side. The spectrometer used was the spectrometer Princeton Instruments SP2150 ($\Delta \lambda =0.4$ nm, $\Delta E =1.7$ meV for $E_{photon}$= 2.28 eV). For the microcavity, the ARPL maps of the position presented in figure \ref{Figure3} a) have been obtained with the same configuration. In this case, the focus and collection were done on the silver side.}\\
\textit{A second configuration, with the same spectrometer was used for the ARPL maps on figure \ref{Figure3} b) and c): the focus was made by a 5-cm lens through the Bragg mirror and the collection with the 0.9 NA objective from the silver side.} \\
\textit{In the case of the PL spectroscopy of the microcavity, in figure \ref{Figure4} and in the section 5 of the supplementary, a third configuration was opted: a 0.6 NA objective was used for focusing the pump laser and collecting the PL from the silver side. In this case, the spectrometer was the Princeton Instruments SP2500 ($\Delta \lambda =0.09$ nm, $\Delta E =0.38$ meV for $E_{photon}$= 2.28 eV).}\\

\noindent \textbf{Theory of the strong coupling regime}
\textit{The strong coupling between an exciton and the photonic mode of a microcavity can be simplified into a two-level model. The system Hamiltonian in the two dimensional basis, ($|X>$,$|ph>$), can be then written as such : }

\vspace*{-5pt}

\begin{equation*}
H= 
\begin{pmatrix}
E_{ph}(\theta) -i\gamma_{ph} & V \\
V  & E_{X} -i\gamma_{X}
\end{pmatrix},
\end{equation*} 

\noindent \textit{where V is the coupling strength, $\gamma_{X}$ and $\gamma_{ph}$ are the excitonic and photonic linewidths. The photonic mode dispersion is given by :}

\vspace*{-5pt}

\begin{equation*}
\centering
E_{ph}(\theta)=(E_X+\delta)/\sqrt{1-(sin(\theta)/n_{eff})^2},
\end{equation*} 

\noindent \textit{with $n_{eff}$ the microcavity effective refractive index and $\delta=E_0-E_X$ the detuning which is the difference between the excitonic energy, $E_X$, and the photonic mode energy at normal incidence $E_0$.} \\

\textit{Solving the system's Schr{\"o}dinger equation gives rise to two new quasi-particles : the lower and upper cavity polaritons, which are the coherent superposition of the photonic and excitonic states. The eingenvalues of the lower and upper cavity polaritons, $\mu_{LP}(\theta)$ and $\mu_{UP}(\theta)$ are given by:}

\vspace*{-5pt}

\begin{equation*}
\centering
\begin{split}
\mu_{UP,LP}(\theta)&=\frac{1}{2}[E_{ph}(\theta)+E_{X}-i(\gamma_{ph}+\gamma_{X})]\\
& \pm\sqrt{V^2+\frac{1}{4}[E_{X}-E_{ph}(\theta)+i(\gamma_{ph}-\gamma_{X})]^2},
\end{split}
\end{equation*} 

\vspace*{5pt}

\noindent \textit{where the real parts of $\mu_{UP,LP}$ correspond to the cavity polariton energy dispersions, $E_{LP}(\theta)$ and $E_{UP}(\theta)$, and the imaginary part to the cavity polariton linewidths, $\gamma_{LP}(\theta)$and $\gamma_{UP}(\theta)$.}\\


\noindent \textbf{Estimation of the polariton propagation length}

\textit{The polariton propagation length is given by $L=v_{LPB}\times t_{LPB}$, where $t_{LPB}$ is the polariton lifetime and $v_{LPB}$ is the polariton velocity. The polariton lifetime can be approximated to be close to the photonic mode lifetime $t_{ph}=\hbar Q/2E_0$, with $\hbar$ the reduced Planck constant, $Q$ the quality factor, and $E_0$ the photonic mode energy at normal incidence. In our case $Q=92$ \cite{Bouteyre2019}, which gives an estimation of the polariton lifetime $t_{LPB}\approx 13$ fs. In the case of a propagation of a polariton condensate, the polariton velocity is given by $v_{LPB}=(\hbar\times k_{prop})/m_{LPB}$, where $m_{LPB}$ is the lower polariton effective mass, and $k_{prop}$ is the wavenumber of the propagated polariton condensate \cite{Wouters2008}. The polariton effective mass, $m_{LPB}$, is related to the curvature of the LPB branch at $k_{//}=0$ and can be obtained with $m_{LPB}=\hbar^2/(2\times C)$, where C is the coefficient of a parabola ($E_p=E_0^p+Ck_{//}^2$) fitting the LPB at low $k_{//}$. Using the parameters found to fit the LPB dispersion in figure \ref{Figure3} a) (figure \ref{Figure3} b)), the LPB effective mass is found to be 8.3(8.9) $eV.ps^2.\mu m^{-2}$. The wavenumber of the propagated polariton condensate is related to the emission angle, $\theta_{BEC}$, and photon energy, $E_{BEC}$, of the polariton condensate as $k_{prop}=(E_{BEC}/c\hbar)\sin(\theta_{BEC})$. In our case, $\theta_{BEC}$ and $E_{BEC}$  correspond to the lasing emission angle and energy: $E_{lasing}\approx 2.28$ eV and $\theta_{lasing}$= 22.4$^\circ$ and 15.8$^\circ$ for the figures \ref{Figure3} a) and b), respectively. The polariton propagation lengths of 470 nm and 309 nm are found respectively for the figures \ref{Figure3} a) and b).}  \\

\noindent \textbf{Study of the free spectral range of the lasing spectra to determine the characteristic size of the MAPB pseudo-cavities}

\textit{The method used to obtain the characteristic size of the MAPB pseudo-cavities is the following. The free spectral range in energy, $FSR_E=E_{m+1}-E_{m}$ with m an integer, of the lasing spectra is studied by collecting the average energies of the modes shown in figure \ref{Figure4} c). The difference in energy between the $m^{th}$ modes and the $1^{st}$ mode is plotted against the modes numbers and is fitted with a linear function. The cavity length is retrieved from $L=hc/(n s)$, where $n=2.3$ is the MAPB refractive index, $c$ is the light velocity and $s$ is the slope of the linear function. }



\section*{Funding Information}

This work is supported by Agence Nationale de la Recherche (ANR) within the projects POPEYE (ANR-17-CE24-0020) and EMIPERO (ANR-18-CE24-0016). The work of P.Bouteyre is supported by the Direction G\'en\'erale de l$^\prime$Armement (DGA).

\section*{Acknowledgments}

We thank Rasta Ghasemi, from Institut d$^\prime$Alembert de l$^\prime$\'Ecole Normale Sup\'erieure de Paris-Saclay, for her technical support in the evaporation of the metallic mirrors. 

\section*{Supplemental Documents}

The following files are available free of charge.
\begin{itemize}
  \item Additional information (PDF) on the fabrication of the MAPB/PMMA sample and the 3$\lambda/2$ MAPB-based microcavity, the PL spectroscopy carried on the MAPB/PMMA sample to demonstrate the random lasing, additional data on the ARPL of the MAPB/PMMA sample and the three positions of the 3$\lambda$/2 MAPB based-microcavity, the fitting method of the lower polariton dispersion curves of the three positions in figure 3, slices taken at given angles and at given energies of the ARPL maps in figure 3, PL spectra as a function of the pump power on another position of the microcavity, a study of the impact of the quality factor on the emission divergence and the sketch of the excitation configurations of the MAPB/PMMA sample and the 3$\lambda/2$ MAPB-based microcavity. 
\vspace*{5pt}
\end{itemize}

\bibliography{mybib}
\bibliographystyle{ieeetr}


\end{document}


\title{Supplementary for :  \vspace{0.5cm} \\ Directing random lasing emission using cavity exciton-polaritons}

\author{P. Bouteyre}
\affiliation{Universit\'e Paris-Saclay, ENS Paris-Saclay, CNRS, Centrale Supelec, LuMIn, 91405 Orsay, France}
\author{H. S. Nguyen}
\affiliation{Universit\'e de Lyon, Insitut des Nanotechnologies de Lyon - INL, UMR CNRS 5270, CNRS, Ecole Centrale de Lyon, Ecully, F-69134, France}
\author{J.-S. Lauret}
\affiliation{Universit\'e Paris-Saclay, ENS Paris-Saclay, CNRS, Centrale Supelec, LuMIn, 91405 Orsay, France}
\author{G. Tripp\'e-Allard}
\affiliation{Universit\'e Paris-Saclay, ENS Paris-Saclay, CNRS, Centrale Supelec, LuMIn, 91405 Orsay, France}
\author{G. Delport}
\affiliation{Universit\'e Paris-Saclay, ENS Paris-Saclay, CNRS, Centrale Supelec, LuMIn, 91405 Orsay, France}
\author{F. L\'ed\'ee}
\affiliation{Universit\'e Paris-Saclay, ENS Paris-Saclay, CNRS, Centrale Supelec, LuMIn, 91405 Orsay, France}
\author{H. Diab}
\affiliation{Universit\'e Paris-Saclay, ENS Paris-Saclay, CNRS, Centrale Supelec, LuMIn, 91405 Orsay, France}
\author{A. Belarouci}
\affiliation{Universit\'e de Lyon, Insitut des Nanotechnologies de Lyon - INL, UMR CNRS 5270, CNRS, Ecole Centrale de Lyon, Ecully, F-69134, France}
\author{C. Seassal}
\affiliation{Universit\'e de Lyon, Insitut des Nanotechnologies de Lyon - INL, UMR CNRS 5270, CNRS, Ecole Centrale de Lyon, Ecully, F-69134, France}
\author{D. Garrot}
\affiliation{Groupe d$^\prime$Etude de la Mati\`ere Condens\'ee, Universit\'e de Versailles Saint Quentin En
Yvelines, Universit\'e Paris-Saclay, 45 Avenue des Etats-Unis, 78035, Versailles, France}
\author{F. Bretenaker}
\affiliation{Universit\'e Paris-Saclay, ENS Paris-Saclay, CNRS, Centrale Supelec, LuMIn, 91405 Orsay, France}
\author{E. Deleporte }
\email{emmanuelle.deleporte@ens-paris-saclay.fr}
\affiliation{Universit\'e Paris-Saclay, ENS Paris-Saclay, CNRS, Centrale Supelec, LuMIn, 91405 Orsay, France}


%
%
%




\renewcommand{\theequation}{S\arabic{equation}}
\renewcommand{\thefigure}{S\arabic{figure}}



\maketitle

\section{Sample fabrication methods}

See the methods section of our previous article \cite{Bouteyre2019} for the microcavity's fabrication. The same fabrication method is used for the MAPB/PMMA sample except that no silver layer has been evaporated on top of the sample and the substrate used was a 2.6x2.6 $cm^2$ quartz substrate instead of the commercial Bragg mirror.   


\vspace*{20pt}

\section{Random lasing of the MAPB/PMMA sample}

\begin{figure*}[h!]
\centering
  \includegraphics[width=\linewidth]{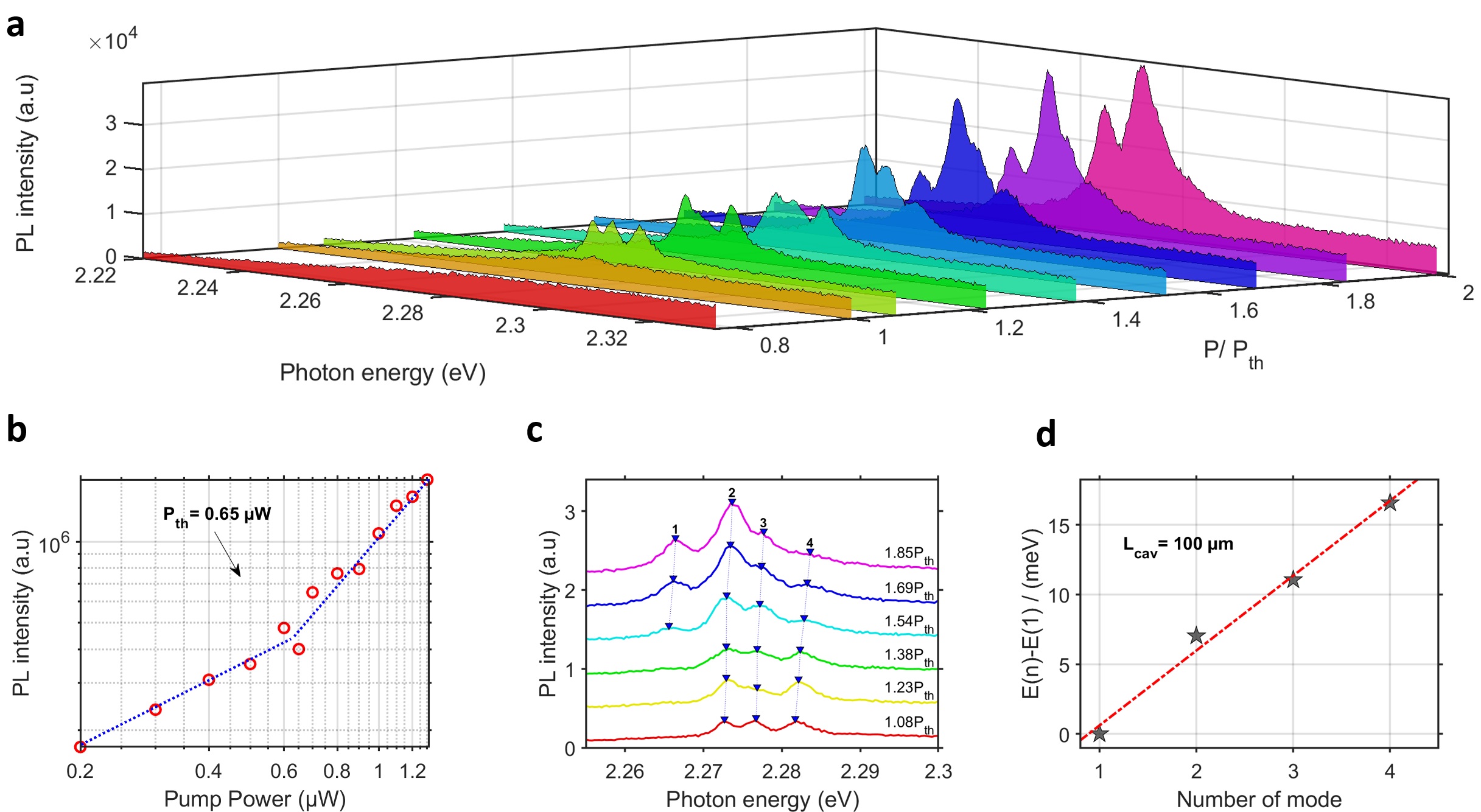}
  \caption{Random lasing action of the MAPB/PMMA sample. The resolution is here of 1.7 meV for the energy axis. \textbf{a)} PL spectra of the MAPB/PMMA sample at different pumping powers. \textbf{b)} Log-log PL intensity spectrum as a function of the pumping powers. \textbf{c)} Five PL spectra above the lasing threshold of a) are plotted with a vertical offset for better readability. Four lasing modes are indicated by blue triangles and the blue dotted lines are guides for the eye. \textbf{d)} Study of the free spectral range in energy, $FSR_E=E_{m+1}-E_{m}$ with $m$ an integer, of the lasing spectra of c) to obtain the characteristic size of the MAPB pseudo-cavities. The difference in energy between the $m^{th}$ modes and the $1^{st}$ mode is plotted against the modes numbers and is fitted with a linear function. The cavity length is retrieved from $L=hc/(n s)$, where $n=2.3$ is the MAPB refractive index, $c$ is the light velocity and $s$ is the slope of the linear function.} 
 \label{FigureS1}
\end{figure*}

Figure \ref{FigureS1} presents the PL spectroscopy results demonstrating the random lasing from the MAPB/PMMA sample. The pump laser was focused, and the PL collected, by a 0.9 NA objective on the PMMA side. Figure \ref{FigureS1} a) shows the photoluminescence spectra of the MAPB/PMMA sample with the pump power varying from 0.8 to 2 times the threshold power ($P_{th}$). Below the threshold, the photoluminescence spectrum is similar to the PL of the bare perovskite presented in Figure 1 b) of the article with an FWHM of around 100 meV. When the pump power is increased, and when it reaches the threshold, three small peaks of FWHM of around 3 meV appear on top of the PL spectrum. As the pump power increases the peaks gain in intensity and new peaks appear. A broadening of the overall lasing signal occurs at higher pumping powers due to the increasing number of modes overlapping because of their spectral proximity. At 2 $P_{th}$, only two modes are well defined. The figure \ref{FigureS1} b) shows the log-log integrated PL as a function of the pump power and gives a pump threshold of 0.65 $\mu$W. The integrated PL was taken between 2.26 eV and 2.29 eV to take into account the contribution of all of the modes. Five PL spectra above the lasing threshold of figure \ref{FigureS1} a) are plotted in figure \ref{FigureS1} c) with a vertical offset for better readability. Four lasing modes are indicated by blue triangles and the blue dotted lines are guides for the eye. A slight blue-shift of the lasing modes can be observed with the pumping power increasing, which is probably caused by the perovskite thermal effects.\\

This random lasing is coherent with the roughness of the perovskite layer presented in the AFM scan image in figure 1 c) of the article : the grains of 500 nm to 1 $\mu$m of size provides the necessary light scattering for the random lasing action. The grains of perovskite scatters the light, and the scattering is strong enough to get some light to scattered back to its original position. The signal collected by the microscope objective is the lasing emission scattered out of the sample plane. Moreover, the perovskite of refractive index around 2.4 is sandwiched between the quartz substrate and the layer of PMMA of refractive index around 1.5. This creates a waveguide which confines the light within the plane. The laser modes lie between the energies 2.26 eV and 2.29 eV. This energy range corresponds to the range where the ASE (Amplified Spontaneous Emission) takes place in MAPB thin films \cite{cadelano2015can} where the net gain is the highest. While the light is scattered in different ways within the surface of the perovskite, only the loops whose characteristic size enables modes in this region of energy participate in the random lasing. 

\newpage

Figure \ref{FigureS1} d) shows the study of the free spectral range in energy, $FSR_E=E_{m+1}-E_{m}$ with $m$ an integer, of the lasing spectra of figure \ref{FigureS1} c) to obtain the characteristic size of the MAPB pseudo-cavities. The difference in energy between the $m^{th}$ modes and the $1^{st}$ mode is plotted against the modes numbers and is fitted with a linear function. The modes energies corresponds to the average energies of the modes shown in figure \ref{FigureS1} c). The cavity length is retrieved from $L=hc/(n s)$, where $n=2.3$ is the MAPB refractive index, $c$ is the light velocity and $s$ is the slope of the linear function. For planar cavities, a factor of 2 would appear in the denominator as the round trip must be taken into account, here the cavity length represents the perimeter of the optical loop. This study gives an average characteristic size of the MAPB pseudo-cavities of around 100 $\mu$m. This size is much larger than the thickness of the sample (of around 500 nm) and confirms that the random lasing occurs in the sample plane.

\vspace*{40pt}

\newpage

\section{Additional data of the ARPL maps in figures 2 and 3}

\vspace*{10pt}

\subsection{Additional data for the ARPL maps of the MAPB/PMMA sample in figure 2}

\vspace*{10pt}

\begin{figure*}[h!]
\centering
  \includegraphics[width=\linewidth]{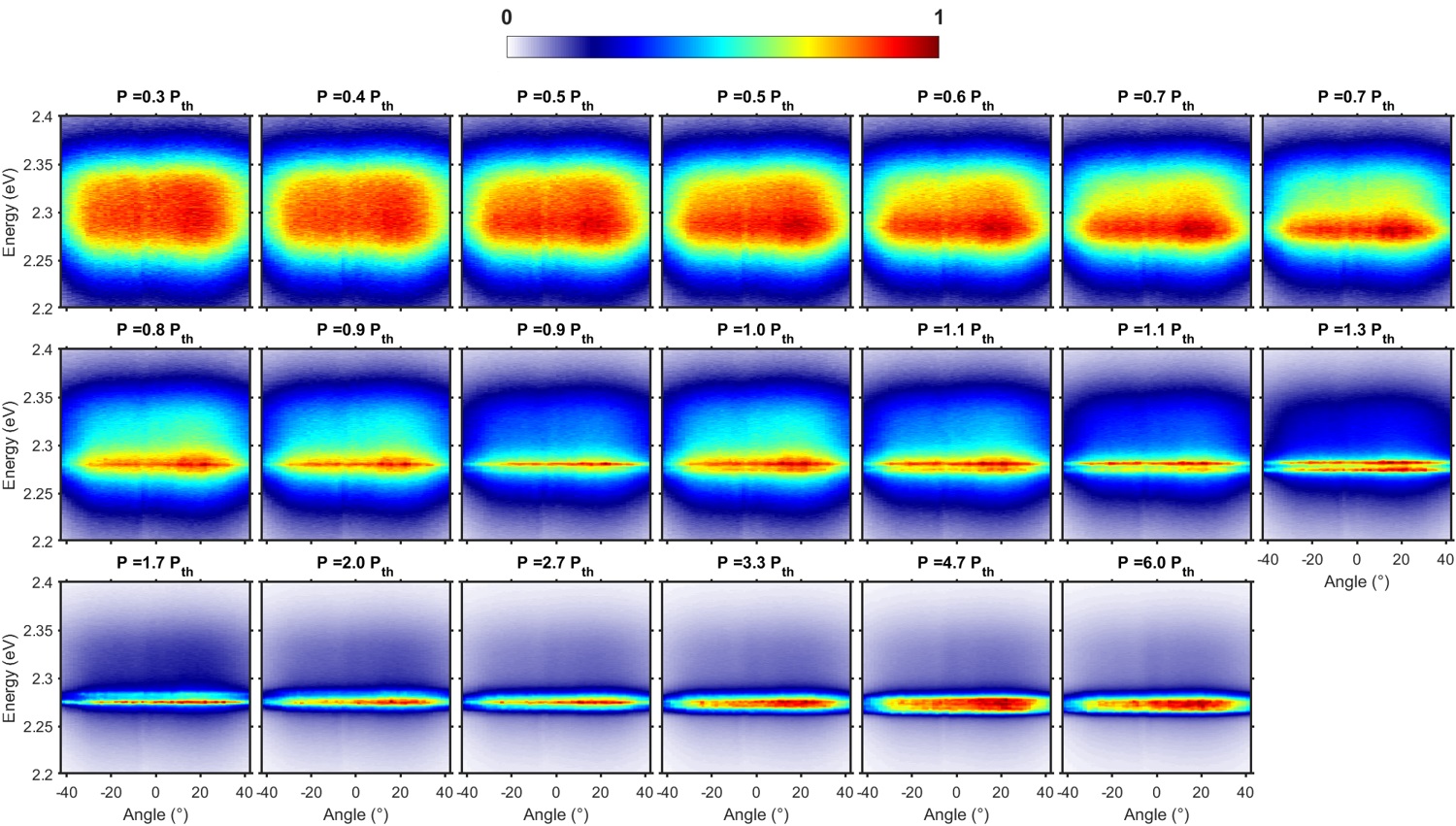}
  \caption{Angle-resolved photoluminescence maps at different pumping powers (from 0.3 to 6.0 $P_{th}$) of the ARPL maps of the MAPB/PMMA sample in figure 2.} 
 \label{FigureS2}
\end{figure*}

\begin{figure*}[h!]
\centering
  \includegraphics[width=0.8\linewidth]{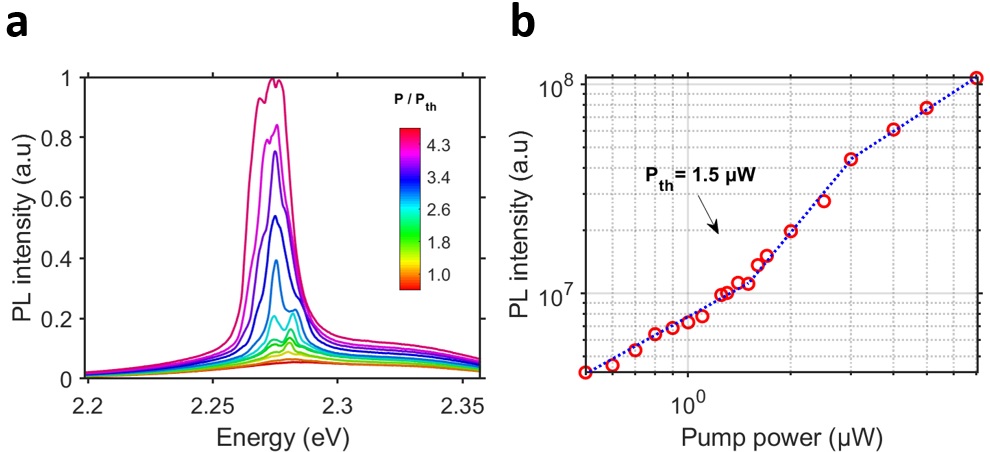}
  \caption{Angle-integrated PL of the ARPL maps of the MAPB/PMMA sample in figure 2. \textbf{a)} Angle-integrated PL spectra at different pumping powers. \textbf{b)} Curve of the integrated PL intensity as a function of the pumping power.} 
 \label{FigureS3}
\end{figure*}

\newpage

\subsection{Additional data for the ARPL maps of the 3$\lambda$/2 MAPB-based microcavity in figure 3a)}

\vspace*{30pt}

\begin{figure*}[h!]
\centering
  \includegraphics[width=\linewidth]{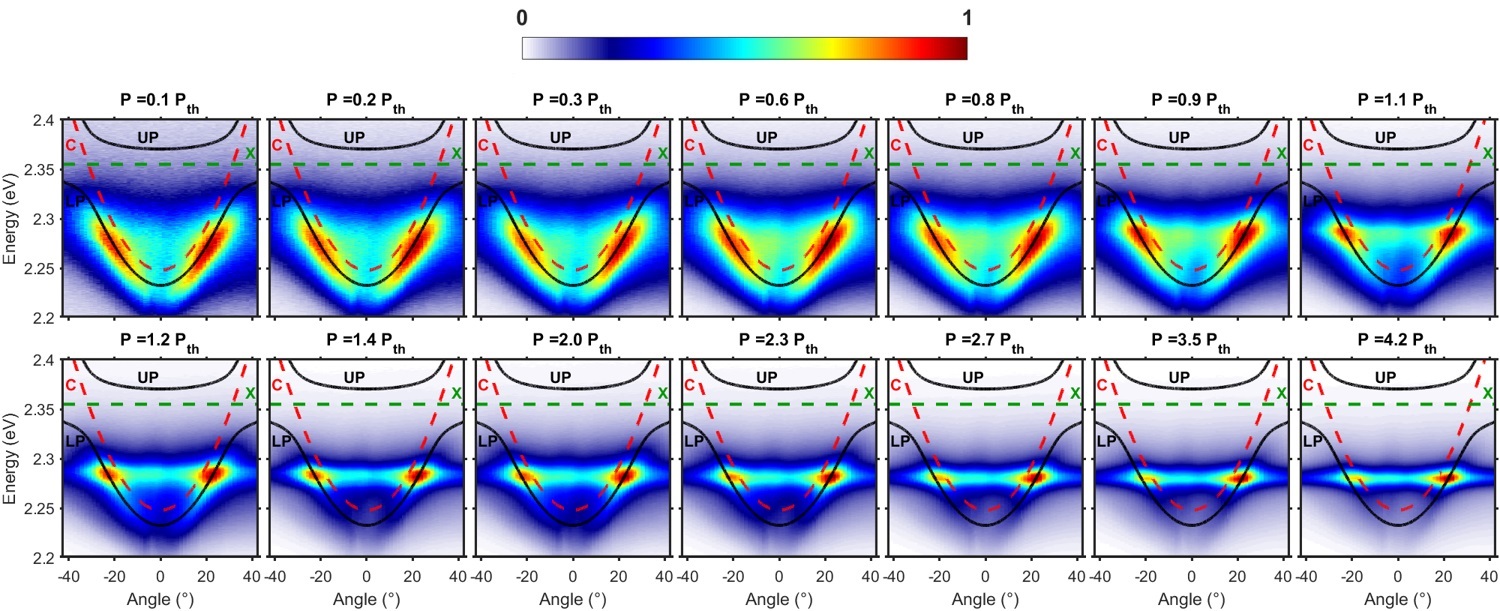}
  \caption{Angle-resolved photoluminescence maps at different pumping powers (from 0.1 to 4.2 $P_{th}$) of the ARPL maps of the 3$\lambda$/2 MAPB-based microcavity in figure 3a).} 
 \label{FigureS4}
\end{figure*}

\begin{figure*}[h!]
\centering
  \includegraphics[width=0.8\linewidth]{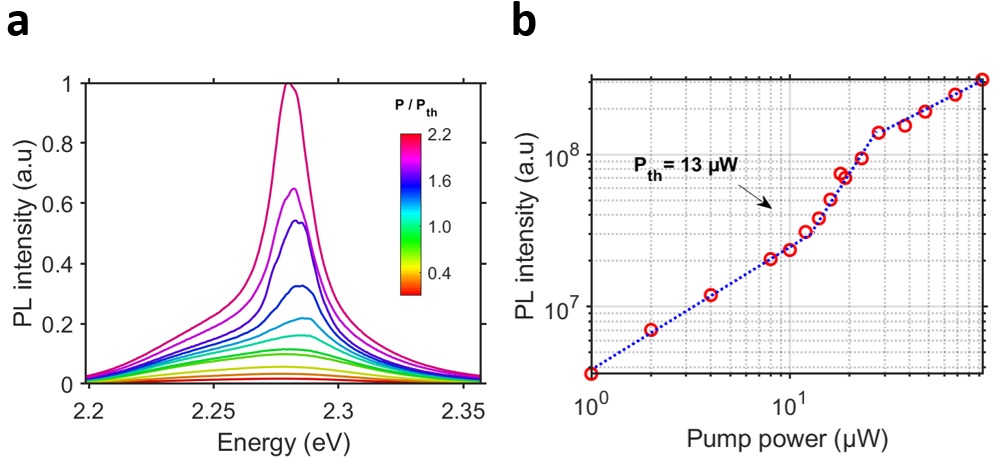}
  \caption{Angle-integrated PL of the ARPL maps of the 3$\lambda$/2 MAPB-based microcavity in figure 3a). \textbf{a)} Angle-integrated PL spectra at different pumping powers. \textbf{b)} Curve of the integrated PL intensity as a function of the pumping power.} 
 \label{FigureS5}
\end{figure*}

\newpage

\subsection{Additional data for the ARPL maps of the 3$\lambda$/2 MAPB-based microcavity in figure 3b)}

\vspace*{30pt}

\begin{figure*}[h!]
\centering
  \includegraphics[width=\linewidth]{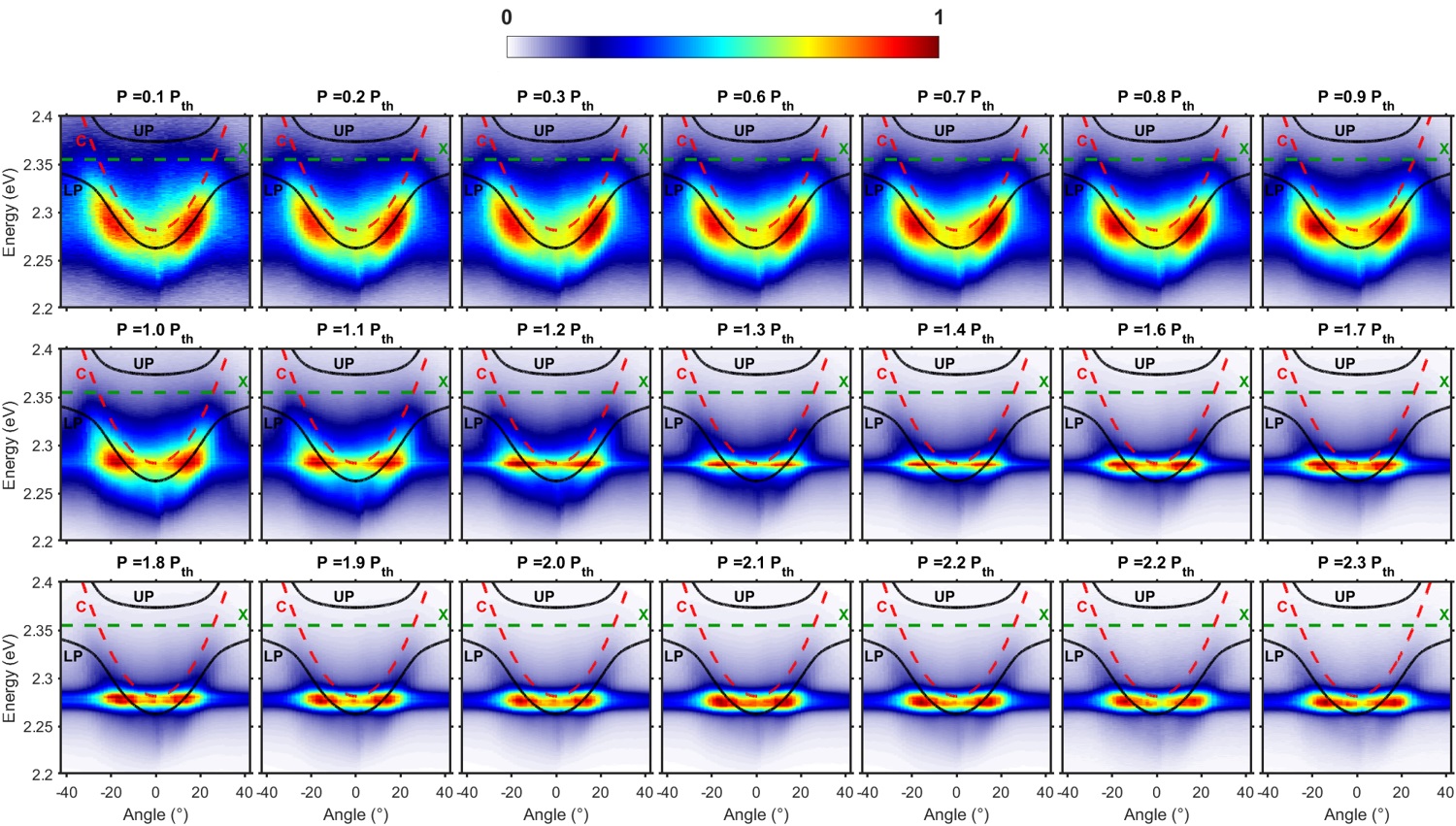}
  \caption{Angle-resolved photoluminescence maps at different pumping powers (from 0.1 to 2.3 $P_{th}$) of the ARPL maps of the 3$\lambda$/2 MAPB-based microcavity in figure 3b).} 
 \label{FigureS6}
\end{figure*}

\begin{figure*}[h!]
\centering
  \includegraphics[width=0.8\linewidth]{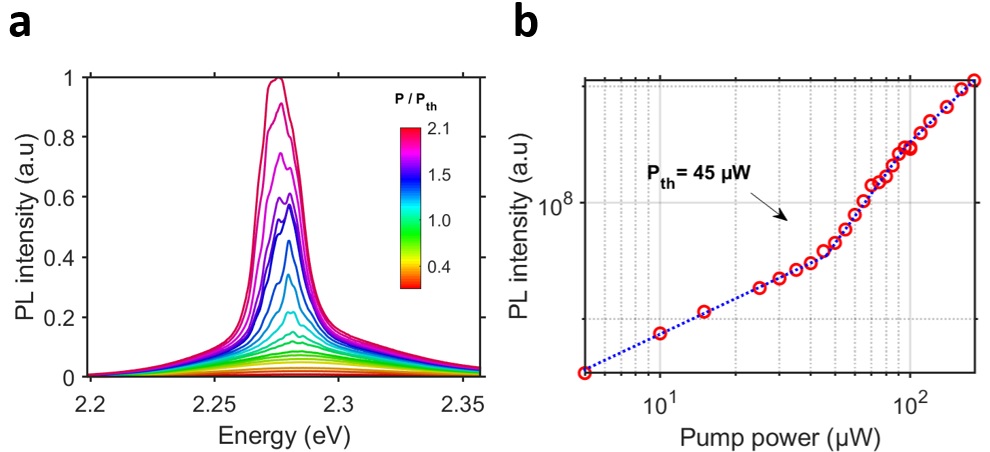}
  \caption{Angle-integrated PL of the ARPL maps of the 3$\lambda$/2 MAPB-based microcavity in figure 3b). \textbf{a)} Angle-integrated PL spectra at different pumping powers. \textbf{b)} Curve of the integrated PL intensity as a function of the pumping power.} 
 \label{FigureS7}
\end{figure*}

\newpage

\subsection{Additional data for the ARPL maps of the 3$\lambda$/2 MAPB-based microcavity in figure 3c)}

\vspace*{30pt}

\begin{figure*}[h!]
\centering
  \includegraphics[width=\linewidth]{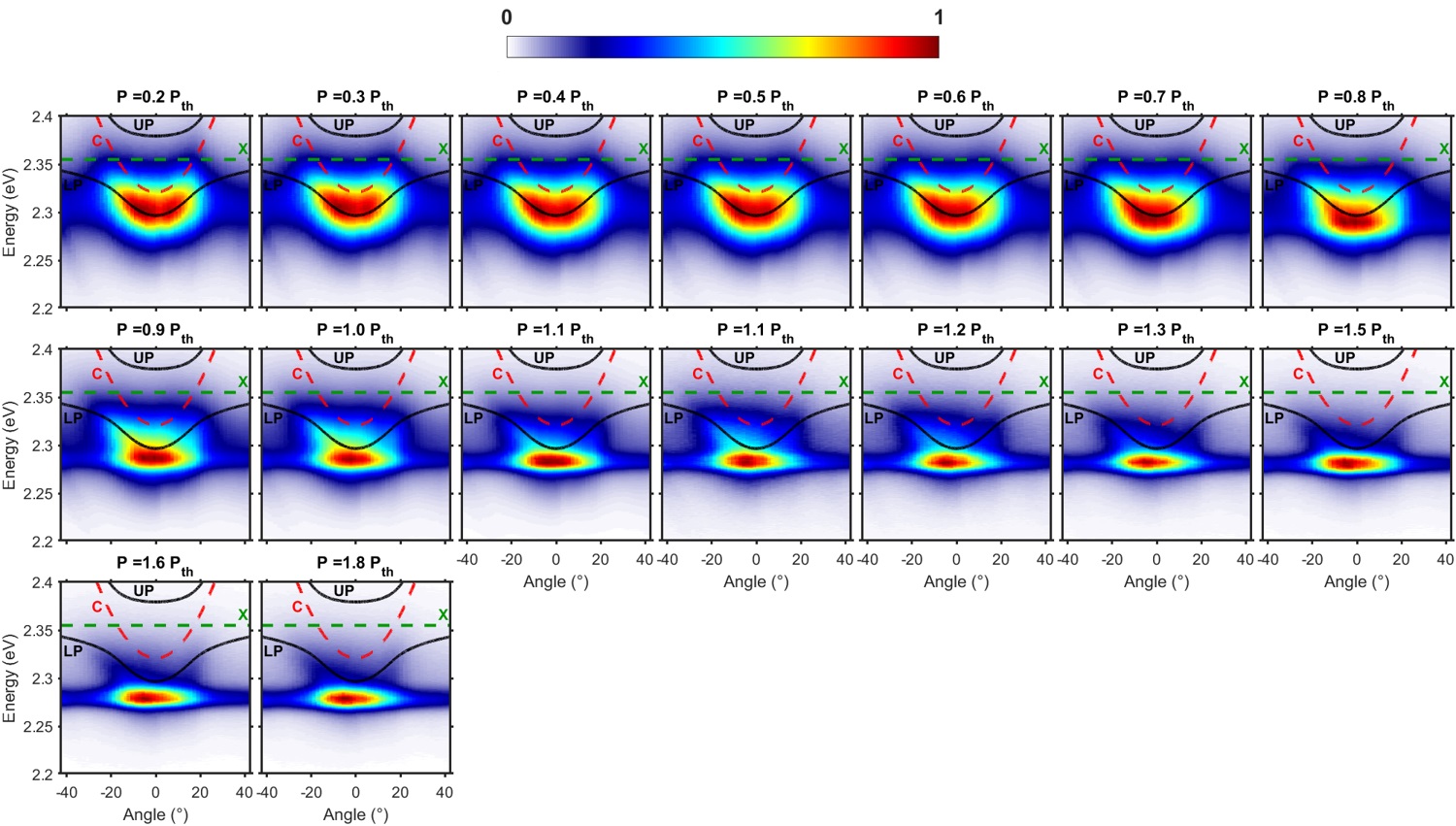}
  \caption{Angle-resolved photoluminescence maps at different pumping powers (from 0.2 to 1.8 $P_{th}$) of the ARPL maps of the 3$\lambda$/2 MAPB-based microcavity in figure 3c).} 
 \label{FigureS8}
\end{figure*}

\begin{figure*}[h!]
\centering
  \includegraphics[width=0.8\linewidth]{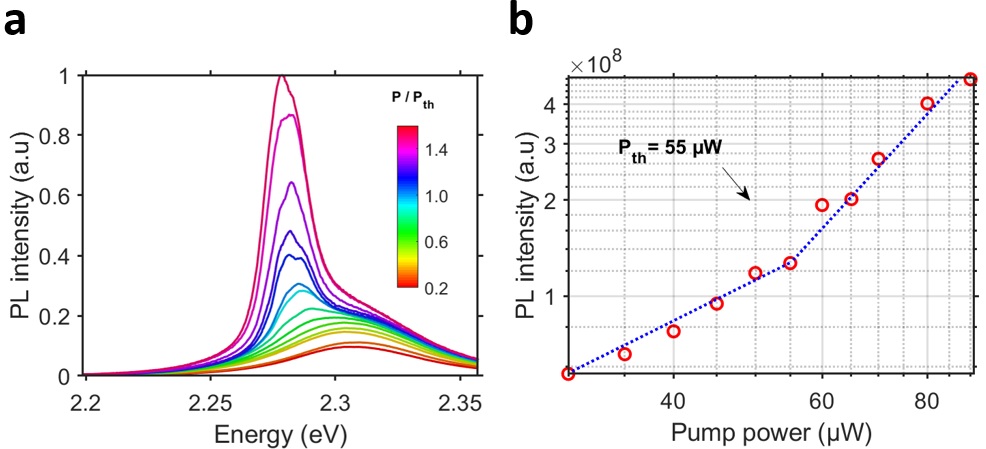}
  \caption{Angle-integrated PL of the ARPL maps of the 3$\lambda$/2 MAPB-based microcavity in figure 3c). \textbf{a)} Angle-integrated PL spectra at different pumping powers. \textbf{b)} Curve of the integrated PL intensity as a function of the pumping power.} 
 \label{FigureS9}
\end{figure*}

\newpage

\section{Fitting method of the lower polariton dispersion curves of the ARPL maps in figure 3}

\vspace*{10pt}

\begin{figure*}[h!]
\centering
  \includegraphics[width=0.8\linewidth]{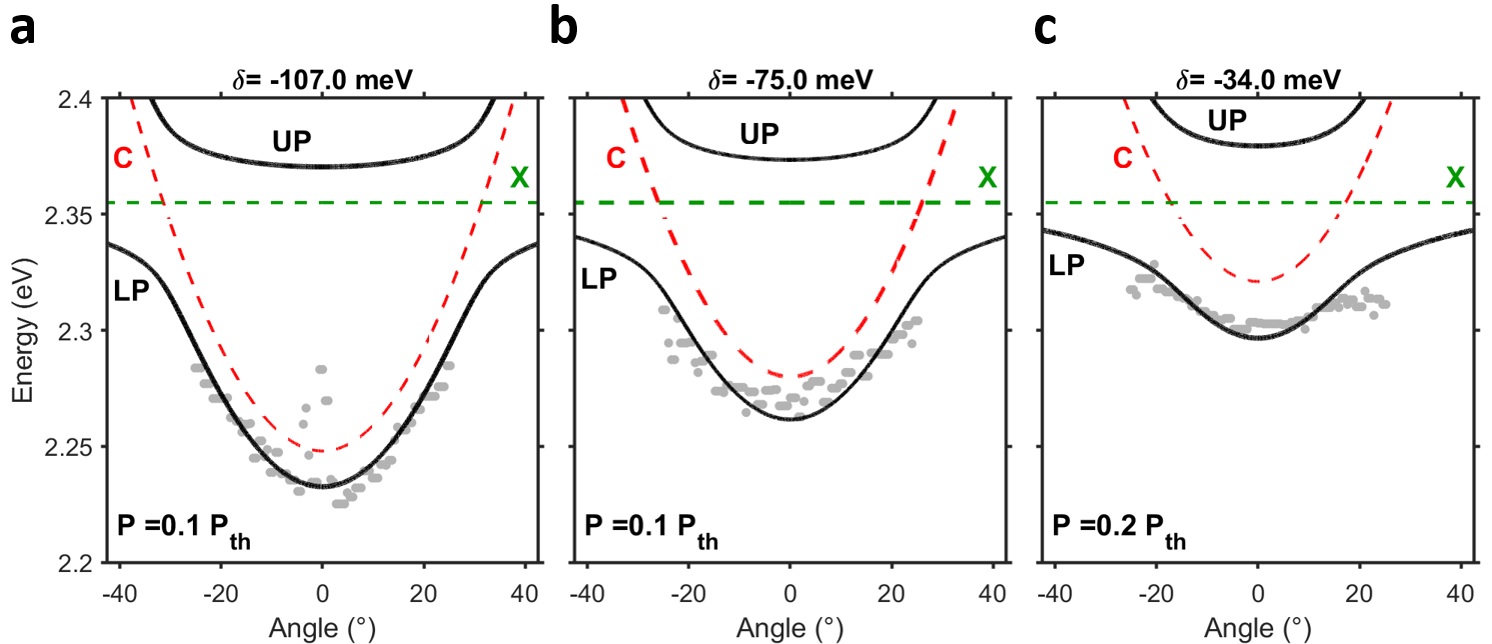}
  \caption{a) to d) Fitting of the lower polariton dispersion curves of the three positions in figures 3 a-(i), b-(i), c-(i), respectively. The grey dots represent the experimental data from the Lorentzian fits of the slices at different angles. The fitted polariton dispersions are plotted in black, the uncoupled excitonic energy in green and the uncoupled photonic mode in red.} 
 \label{FigureS10}
\end{figure*}

This supplementary section shows the fitting method of the lower polariton dispersion curves of the three positions in figures 3 a-(i), b-(i), and c-(i). The experimental dispersion curves are obtained by fitting slices of the photoluminescence maps (between -20 $^\circ$ and 20 $^\circ$) with a Lorentzian function. The Lorentzian centers are reported in energy/angle diagrams shown in figure \ref{FigureS10}. The collected dispersions curves are then fitted to the polariton two-level model presented in the method section using the parameters found in our previous study \cite{Bouteyre2019} ($n_{eff}$=1.75, $E_X=2.355$ eV, g=48.7 meV, $\gamma_X$=90 meV and $\gamma_{ph}$=25 meV) and with the detuning, $\delta$, as the only free parameter. 

\vspace*{20pt}

\newpage

\section{Slices at given angles and given energies of the ARPL maps in figure 3}

\vspace*{10pt}

\begin{figure*}[h!]
\centering
  \includegraphics[width=0.8\linewidth]{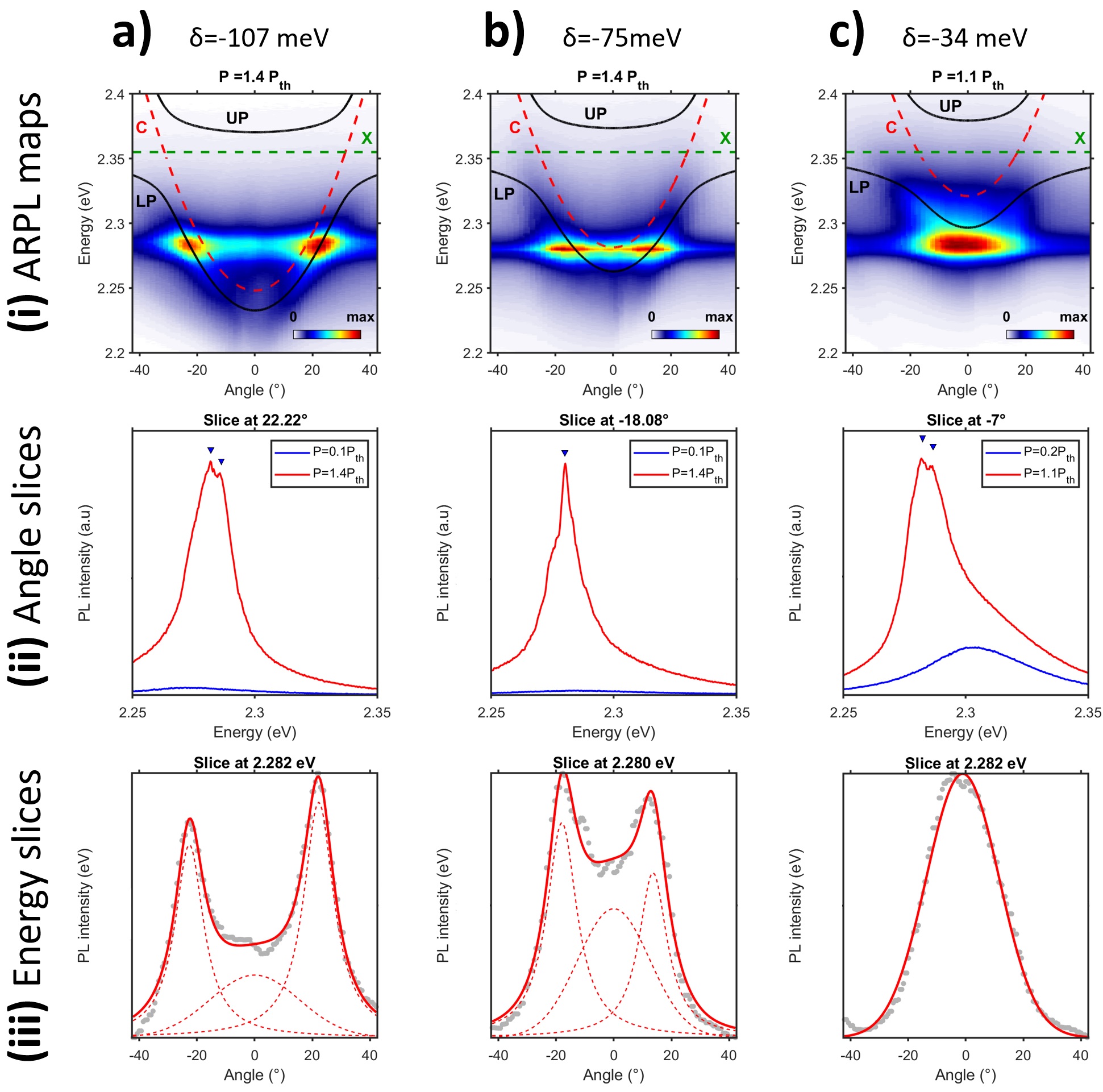}
  \caption{a) to c) Slices at given angles and given energies of the ARPL maps in figures 3 a-ii), b-ii), and c-ii), respectively. The row (i) shows the ARPL maps with the fitted polariton dispersions plotted in black the uncoupled excitonic energy and photonic modes in green and red respectively. The row (ii) shows the slices of the ARPL maps for given angles. The row (iii) shows the slices of the ARPL maps for given energies fitted with Lorentzian and Gaussian functions. Lorentzian functions fit the lasing signal in a-iii) and b-iii), Gaussian functions fit the background photoluminescence (PL) in a-iii) and b-iii) and the overall PL signal of c-iii).} 
 \label{FigureS11}
\end{figure*}

This supplementary section shows the slices taken at given angles and at given energies of the ARPL maps in figures 3 a-ii), b-ii), and c-ii). In figure \ref{FigureS11}, the ARPL maps above lasing threshold are reproduced in the row (i), the slices taken at given angles (vertical slices) are shown in the row (ii), and the slices taken at given energies (horizontal slices) are shown in the row (iii). The vertical slices of the row (ii) helps to better discern the lasing peaks. The fitting of the horizontal slices in row (iii) with Lorentzian and Gaussian functions permits to obtain the values of the lasing peaks angles and divergences. Lorentzian functions fit the lasing signal in figure \ref{FigureS11} a-iii) and b-iii), Gaussian functions fit the background photoluminescence (PL) in figure \ref{FigureS11} a-iii) and b-iii) and the overall PL signal of figure \ref{FigureS11} c-iii).

\vspace*{20pt}

\newpage

\section{Supplementary PL Spectroscopy of another point of the microcavity}

\vspace*{10pt}

\begin{figure*}[h!]
\centering
  \includegraphics[width=\linewidth]{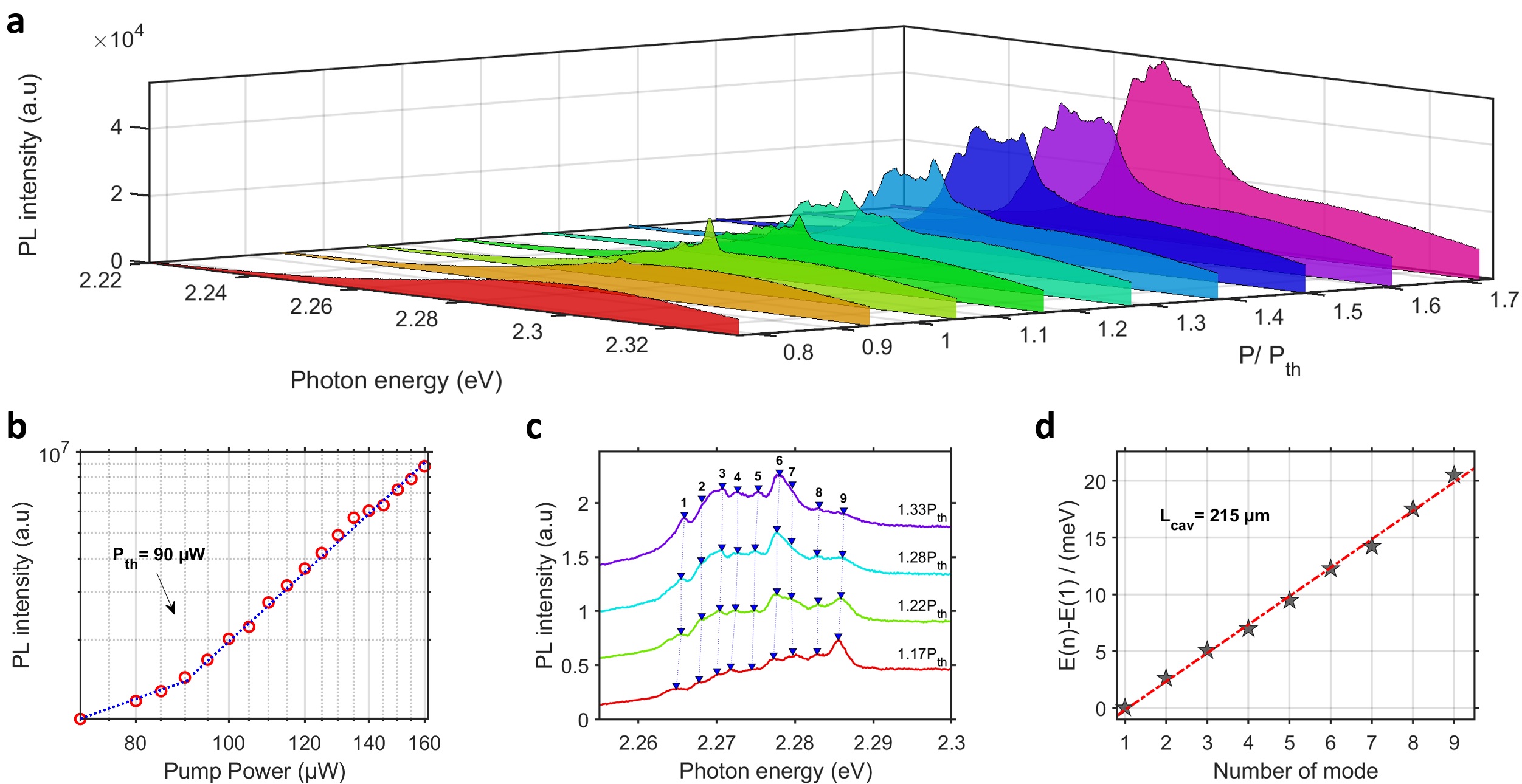}
  \caption{PL spectroscopy of the lasing action from another position of the $3\lambda/2$ microcavity. \textbf{a)} PL spectra of this position at different pumping powers. \textbf{b)} Log-log PL intensity spectrum as a function of the pumping powers. \textbf{c)} Four PL spectra above the lasing threshold of a) are plotted with a vertical offset for better readability. Nine lasing modes are indicated by blue triangles and the blue dotted lines are guides for the eye. \textbf{d)} Study of the free spectral range in energy, $FSR_E=E_{m+1}-E_{m}$ with $m$ an integer, of the lasing spectra of c) to obtain the characteristic size of the MAPB pseudo-cavities. The difference in energy between the $m^{th}$ modes and the $1^{st}$ mode is plotted against the modes numbers and is fitted with a linear function. The cavity length is retrieved from $L=hc/(n s)$, where $n=2.3$ is the MAPB refractive index, $c$ is the light velocity and $s$ is the slope of the linear function.} 
 \label{FigureS12}
\end{figure*}

The figure \ref{FigureS12} a), b) and c) show the photoluminescence spectra, the log-log integrated PL, and a close up of four lasing spectra of another lateral position on the microcavity. Below the threshold, the spectrum is shifted to 2.3 eV with an FWHM of 66 meV due to the coupling of the PL with the microcavity. In this case, the different modes have approximately the same intensity, with a dominant one at 2.285 eV of FWHM of 2 meV for pumping powers close to the threshold. The study of the free spectral range in figure \ref{FigureS12} d) gives an average characteristic size of the MAPB pseudo-cavities of around 215$\mu$m.   \\

\vspace*{20pt}

\newpage
\section{Impact of the quality factor on the divergence}

\vspace*{10pt}

\begin{figure}[h!]
	\centering
		\includegraphics[width=0.66\linewidth]{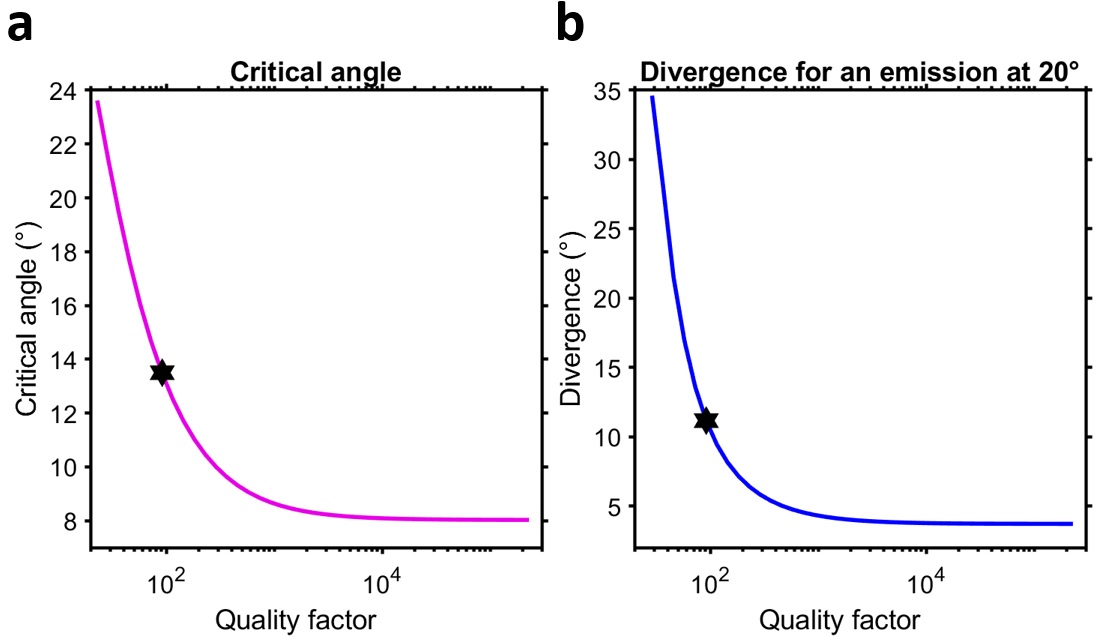}
	\caption{Influence of the cavity quality factor on the a) critical angle under which the emission angle cannot be controlled (in semilog) and b) the divergence of a lasing emission at 20$^\circ$ (in semilog). The black stars indicate the quality factor of around 90 of our case.}
	\label{FigureS13}
\end{figure}

With the numerical approach proposed in the article, we showed that by varying the detuning it is possible to control the emission angle of the microcavity lasing either at normal incidence or between 13.5 $^\circ$ and 60 $^\circ$. However, the divergence of the lasing emission is higher than 10$^\circ$, and the angles between 0 $^\circ$ and 13.5 $^\circ$, called hereafter critical angle, are not accessible. The divergence and critical angle are limited by the lower polariton linewidth, thus by the exciton and the photonic mode linewidths. As the exciton linewidth is intrinsic to the gain material, the only way to decrease the lasing emission divergence and the critical angle is to lower the photonic mode linewidth and thus to improve the microcavity quality factor. Figure \ref{FigureS13} presents the influence of the microcavity quality factor on the critical angle and the divergence of a lasing emission at 20$^\circ$ (in semilog). The black stars indicate the quality factor of around 90 of our case. Both the divergence and the critical angles are reduced with the quality factor increasing. However, the reduction is saturated for quality factors higher than 1000, which means that for a MAPB-based microcavity, a quality factor of 1000 is enough to obtain the best performance of the lasing emission directional filtering. The critical angle cannot be lower than 8$^\circ$ and the divergence of a lasing emission at 20$^\circ$ cannot be lower than 3.7$^\circ$. This is due to the excitonic linewidth, which limits the reduction of the lower polariton linewidth with the quality factor increasing. As a consequence, another way to achieve better filtering would be to use a gain medium exhibiting an exciton with a smaller linewidth.

\vspace*{20pt}

\newpage

\section{Excitation configurations used for the photoluminescence (PL) spectroscopy and angle-resolved photoluminescence (ARPL) measurements}

\begin{figure}[h!]
	\centering
		\includegraphics[width=0.8\linewidth]{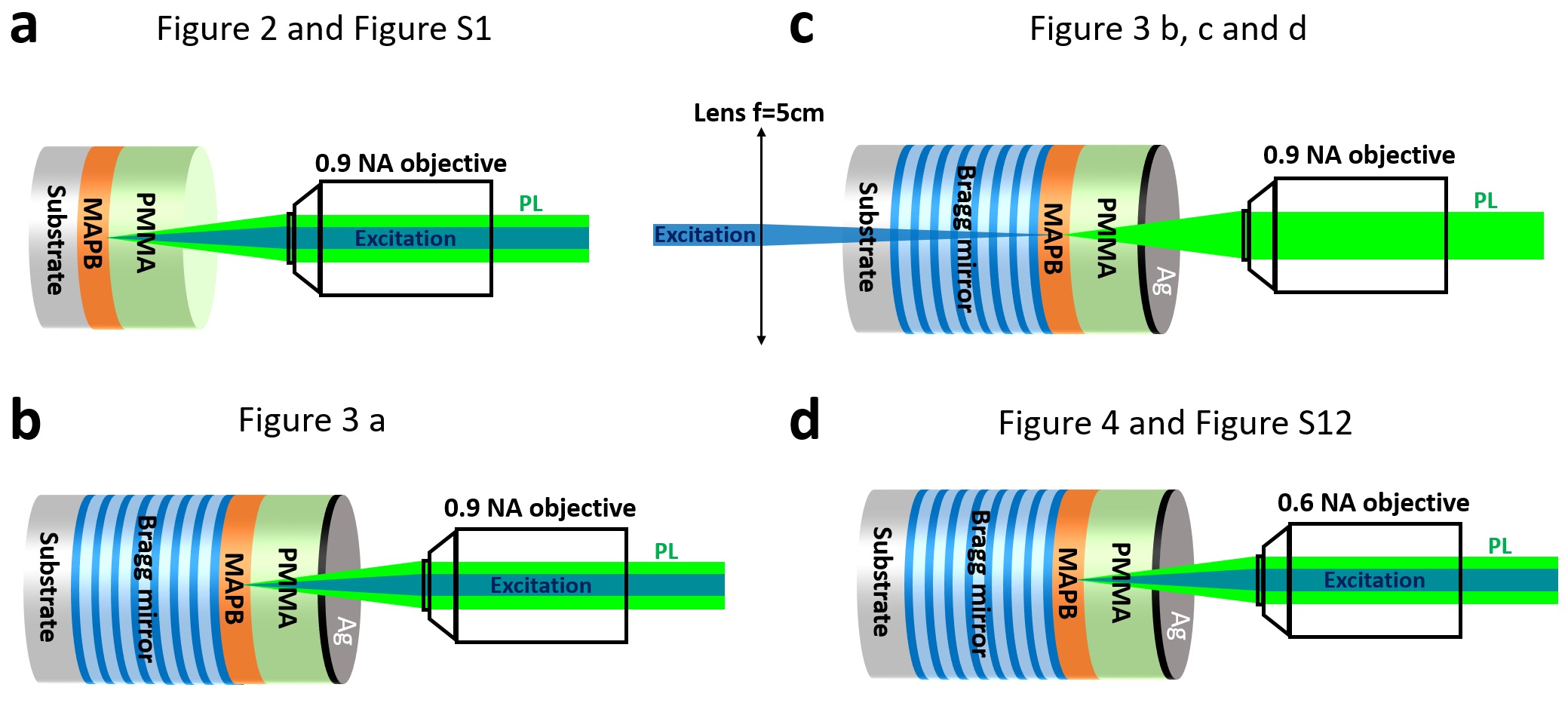}
	\caption{Sketch of the excitation configuration of \textbf{a)} the MAPB/PMMA sample measurements in figure 2 and figure S1 \textbf{b)} the 3$\lambda$/2 MAPB-based microcavity measurements in figure 3 a), \textbf{c)} the 3$\lambda$/2 MAPB-based microcavity measurements in figure 3 b) and c), \textbf{d)} the 3$\lambda$/2 MAPB-based microcavity measurements in figure 4 and figure S12.}
	\label{FigureS14}
\end{figure}

%
%
%
%


\bibliography{mybibsupp}
\bibliographystyle{ieeetr}

